\documentclass[sigconf]{acmart} 

\usepackage{graphicx}
\usepackage{subcaption}
\usepackage{soul}
\AtBeginDocument{%
  \providecommand\BibTeX{{%
    \normalfont B\kern-0.5em{\scshape i\kern-0.25em b}\kern-0.8em\TeX}}}

\setcopyright{acmlicensed}
\copyrightyear{2018}
\acmYear{2018}
\acmDOI{XXXXXXX.XXXXXXX}

\copyrightyear{2024}
\acmYear{2024}
\setcopyright{rightsretained}
\acmConference[DIS '24]{Designing Interactive Systems Conference}{July 1--5, 2024}{IT University of Copenhagen, Denmark}
\acmBooktitle{Designing Interactive Systems Conference (DIS '24), July 1--5, 2024, IT University of Copenhagen, Denmark}\acmDOI{10.1145/3643834.3661572}
\acmISBN{979-8-4007-0583-0/24/07}

\renewcommand{\hl}[1]{#1}

\begin{document}

\title[Sonic Entanglements with Electromyography]{Sonic Entanglements with Electromyography: Between Bodies, Signals, and Representations}

\author{Courtney N. Reed}
\email{c.n.reed@lboro.ac.uk}
\orcid{0000-0003-0893-9277}
\affiliation{%
  \institution{Institute for Digital Technologies,\\Loughborough University London}
  \city{London}
  \country{United Kingdom}
}

\author{Landon Morrison}
\email{l.morrison@imperial.ac.uk}
\orcid{orcid}
\affiliation{%
  \institution{Dyson School of Design Engineering,\\ Imperial College London}
  \city{London}
  \country{United Kingdom}
}

\author{Andrew McPherson}
\email{andrew.mcpherson@imperial.ac.uk}
\orcid{orcid}
\affiliation{%
  \institution{Dyson School of Design Engineering,\\ Imperial College London}
  \city{London}
  \country{United Kingdom}
}
\author{David Fierro}
\email{davidfierro@gmail.com}
\orcid{orcid}
\affiliation{%
  \institution{CICM, Université Paris 8}
  \city{Saint-Denis}
  \country{France}
}

\author{Atau Tanaka}
\email{a.tanaka@gold.ac.uk}
\orcid{0000-0003-2521-1296}
\affiliation{%
  \institution{CSTC, Goldsmiths}
  \city{London}
  \country{United Kingdom;}
}
\affiliation{%
  \institution{Bristol Interaction Group}
  \city{Bristol}
  \country{United Kingdom;}
}
\affiliation{%
  \institution{MSH Paris Nord}
  \city{Saint-Denis}
  \country{France}
}

 \renewcommand{\shortauthors}{Reed, Morrison, McPherson, Fierro, and Tanaka}

\begin{abstract}

This paper investigates sound and music interactions arising from the use of electromyography (EMG) to instrumentalise signals from muscle exertion of the human body. We situate EMG within a family of embodied interaction modalities, where it occupies a middle ground, considered as a ``signal from the inside'' compared with external observations of the body (e.g., motion capture), but also seen as more volitional than neurological states recorded by brain electroencephalogram (EEG). To understand the messiness of gestural interaction afforded by EMG, we revisit the phenomenological turn in HCI, reading Paul Dourish's work on the transparency of ``ready-to-hand'' technologies against the grain of recent posthumanist theories, which offer a performative interpretation of musical entanglements between bodies, signals, and representations. We take music performance as a use case, reporting on the opportunities and constraints posed by EMG in workshop-based studies of vocal, instrumental, and electronic practices. We observe that across our diverse range of musical subjects, they consistently challenged notions of EMG as a transparent tool that directly registered the state of the body, reporting instead that it took on ``present-at-hand'' qualities, defamiliarising the performer's own sense of themselves and reconfiguring their embodied practice.

\end{abstract}





\maketitle

\section{Introduction}

We live in the age of disappearing media, \hl{or so it seems.} 

Apple's Vision Pro headset claims to kick off a ``new era of spatial computing… [that] seamlessly blends digital content with physical space.'' Neuralink's implantable brain-computer interface, suggestively dubbed Telepathy, promises control over ``almost any device, just by thinking.'' Amazon Go corner stores invite shoppers to grab what they need and ``just walk out,'' with zero contact or interaction required for payment.  
 
Each of these technological systems is marked by a rhetoric of transparency, where users are offered (the illusion of) direct control over their experience. \hl{But as media theorists like Wendy Hui Kyong Chun have argued, such a ``notion of transparency has less to do with actual technological operations than with the `microworld' established by computation"} \cite [p. 18]{chun2005software}, \hl{which effectively conceals the workings of technology beneath an invisibility cloak.} If, however, we look beyond the smooth surface of these seamless interactions, we find a messy entanglement of materials, techniques, bodies, languages, and institutions that must come together to sustain the narrative. As a result, the trick of transparency relies on a denial of the underlying technocultural conditions necessary to produce such experiences, trading matters of representation for tropes of total immersion in the essence of being. 

The apparent trend towards disappearing media in the early decades of the twenty-first century provides a backdrop for this paper's focus on the role of Electromyography (EMG) in emergent sonic practices, including its integration with vocal, instrumental, and electroacoustic performance apparatuses. EMG is a technique for recording electric signals associated with the contraction of skeletal muscle tissue and the firing of motor neurons. In HCI applications, it is most commonly performed using a surface electromyogram (sEMG), where conductive electrodes are adhered to the skin in regions above one or more muscles of interest – commonly on the arms, but also \hl{other limbs and elsewhere such as the torso and} vocal musculature. A typical implementation couples these electrodes to a high-gain, low-noise analog amplifier optimised to operate at low AC frequencies, producing one or more continuous-time electrical signals which are often sampled and represented digitally as a time series. As a result of these multiple contingencies, there cannot exist any general or idealised instantiation of EMG. The signal is notionally identified with the neural impulses activating muscular fibres, but the stochastic nature of muscle innervation, combined with the variabilities of cutaneous sensing and the specifics of the analog signal measurement apparatus, mean that every EMG signal will be specific to that combination of body, electronics, signal processing, and context of use.

Because EMG targets bioelectric phenomena that reside somewhere between intention and action, it occupies a somewhat ambiguous position along the continuum of being and representation. On one hand, using ``raw'' biosignals to make music would seem to map sounds to materials in a way that is less prescriptive than a representational format like MIDI (Musical Instrument Digital Interface), which imposes rigid \hl{definitions of} what counts as a musically meaningful parameter and mode of interaction. On the other hand, as we learn from information theorist Geoffrey Bowker, ``raw data is both an oxymoron and a bad idea'' \cite [p. 184]{bowker2008memory}, as the conditions under which data is collected (and deemed worthy of collection), along with the kind of measuring apparatus employed, will invariably influence the end results. The data are ``always already `cooked' and never entirely `raw' '' \cite [p.2]{gitelman2013raw}, as it were. So, how are we to understand the kinds of bioelectric signals collected with EMG---raw or cooked?---and how do they relate to a broader history of meaning-making that underpins musical interactions? Moreover, how do EMG-based musical instruments fit into (or not) the broader rhetoric of transparency and disappearing media, given their interactive modality of contactless performance?

To answer these questions, this paper frames EMG as a material-discursive practice that troubles neat binaries between subjects and objects in musical interactions. We draw together theories from the phenomenological turn in HCI \cite {dourish2001action, gunkel2018relational} and more recent posthumanist approaches to design practice \cite {wakkary2021things, sanches2022diffraction, nicenboim2023decentering}, examining the latter through its articulation to the idea of \textit{entanglement}–a term that is typically invoked to describe a relational network, a thick web of ``matters of being, knowing, and doing, of ontology, of epistemology, and ethics, of fact and value'' \cite [p. 3] {barad2007meeting}. The term has taken hold across several disciplines, ranging the gamut from archaeology \cite {der2016archaeology} to management studies \cite {orlikowski2010sociomateriality}, but a primary touchstone for so-called ``Entanglement HCI'' \cite {frauenberger2019entanglement, morrison_chi2024} has been new materialism and the feminist science studies of scholars like Karen Barad \cite {barad2007meeting} and Donna Haraway \cite {haraway2013situated}. We, too, build on their theories, and in particular, the idea that ``phenomena are produced through agential intra-actions of multiple apparatuses of bodily production,'' and that ``it is through such practices that the differential boundaries between `humans' and `nonhumans,' `culture' and `nature,' the `social' and the `scientific' are constituted'' \cite[p. 817]{barad2003posthumanist}. Thinking in terms of intra-actions, in which there is a co-constitution of elements that make up phenomena, has proven especially useful in trying to make sense of the fuzzy boundaries between bodies, signals, and representations in EMG-based interactions. Indeed, it is often difficult to parse the causal chains of musical material and signal information in EMG setups, as bi-directional flows establish a feedback loop that defamiliarises the performer's own sense of themselves, translating their creative intents and actions. 

This \hl{paper} investigates the messiness of EMG in the case of musical interaction, challenging the rhetoric of transparency by considering what role designers play and what kind of accountability they have in encoding EMG as a set of material-discursive practices. As well, we examine what kind of flexibility performers and listeners have in decoding the apparatus and adapting it across different cultures of use. We begin in the next section by offering some brief historical background for EMG techniques and tracing their affordances as a sensor modality, followed by a discussion of EMG in the context of HCI and artistic practices, with a special focus on challenges presented by the use case of musical performance. We then turn to a series of three case studies, each grounded in work by one of the co-authors, exploring practical implementations of EMG for the body, musical instruments, and vocal settings. These case studies present an opportunity for seeing how EMG can be made to operate according to different modes of interaction, including by directly sonifying the biosignals, parametric gesture-sound mapping, or using machine-learning techniques for training performance systems in real time. Based on auto-ethnographic accounts and case study details provided by the involved co-authors, \hl{the final part of the article offers our theoretical contribution to working with EMG. This is comprised of:}
\begin{itemize}
    \item \hl{a collective discussion on what practical experience from the musical domain might tell us about design interaction with EMG and biodata more broadly, examining data ``tidiness'', representationalism, and transparency;}
    \item \hl{a reframing of EMG that confronts assumptions about it being a direct representation of the body, rather positioning EMG as a boundary-making, entangled practice that is useful in enacting particular views of the body, as well as roles for designers and users; and}
    \item \hl{suggestions for designers, researchers, and artists who want to work with EMG, given this repositioning of EMG as both sensor and actuator }\cite{wobbrock2012seven}\hl{.}
\end{itemize}





\section{Background and Related Work}\label{sec:background}

\subsection{History \& Physiology of Electromyography}\label{sec:emghistory}

EMG has a rather quirky history–it emerges out of electrophysiology research on ``animal electricity'' in the eighteenth century, and in several historical accounts, it is closely tied to Italian physician Luigi Galvani's experiments with dead frogs to demonstrate their muscles twitching as a result of electroshocks \cite{piccolino1997luigi, piccolino1998animal, kazamel2017history, lockhart2017giuditta}. Rather infamously, these experiments sparked a debate between Galvani and Alessandro Volta, with a central point of contention being whether or not the frog legs were acting as conductors of the external stimulus, or whether the electrical signal was emanating from inside the brain. Out of this debate, a whole pseudoscientific practice of ``galvanization'' developed (related to but different from the philosophy of Galvinism), with Galvani's nephew, Giovanni Aldini, attempting to raise the dead through electric shock treatments. Despite not succeeding, his public demonstrations sparked the popular imagination, coming to life in monstrous figures like that of Mary Shelley's \textit{Frankenstein}, where the power to control bioelectric signals holds the key to animating life. Musicologist Sarah Hibberd \cite {hibberd2017good} has noted similarities between this early vitalist discourse and how, around the same time in the press, electricity became a common metaphor for describing music performances, which were said to be ``electrifying to the soul'' and to have the ability to ``mediate between the material body and the invisible mind'' [p. 182]. There is thus a sense in which music \textit{is} bioelectricity and a creative life force, at least in the Romantic imagination of nineteenth-century listeners.

Given its deep entanglements with electrophysiology and vitalist discourses, perhaps it is not surprising that EMG in HCI and music applications today is still discussed in terms of interiority, of direct communication \textit{sans} medium, and of somehow getting closer to the essence of being itself. But taking a closer look at the dense layers of mediation required to collect biosignals using EMG reveals an irreducible imbrication of matter and representation all the way down. 

\hl{In technical terms,} the EMG records bioelectrical activity at the extremity of the central nervous system, such as that of muscle activations involved in the contraction of skeletal muscle tissue. It is a measure of isotonic, isometric, and isokinetic muscle activity generated by the firing of motor neurons \cite{de_luca_derivation_1975}. 
A healthy muscle under relaxation conditions does not display any activity in the EMG signal. By its nature, raw EMG spikes are of random shape, which means one raw burst of EMG activity cannot be precisely reproduced in the same way twice \cite{konrad_abc_2005}. This is linked to the fact that the actual sets of recruited motor units constantly changes during muscle contraction based on their availability. The EMG not a continuous signal, but the sum of discrete neuron impulses. This results in an aperiodic, stochastic signal that poses challenges to information processing. Interactive applications seeking to track participant's state or gesture typically perform signal analysis and feature extraction \cite{phinyomark_feature_2018}. \hl{We describe technical aspects of the EMG in detail in} \cite{tanaka_cmj2023}.

\subsection{The EMG as a Sensor Modality}\label{sec:sensormodality}

In an HCI context, the use of ``EMG'' often denotes the technical interface for eliciting signals, particularly interfaces for sensing through the surface of the skin (surface electromyography, or sEMG). These technical systems are typically framed as a \hl{sensing of existing} neural impulses within the body, which in turn offer a window into a user's intention \cite{10.1186/1475-925x-6-45}. Hence, by sensing EMG signals, the system designer gains a privileged window into intention detached from visible action. We will revisit and \hl{problematise} these premises in Section \ref{sec:discussion}.

As sensor interfaces, EMG sits in an interesting intermediate space between exterior (objective) and interior (subjective) perspectives on the lived body. On the exterior side, consider the contrast with technologies for sensing physical movement. Optical motion capture presents a detached Cartesian view of the movements of points with respect to a static frame of reference, while inertial sensors, such as accelerometers and gyroscopes, offer a device-centric account of the forces and rotations acting on them. In interaction design practice, motion capture and inertial measurement units (IMUs) have complementary strengths and weaknesses \cite{medeiros2014comprehensive}, but both offer an apparent tidiness stemming from their orthogonal Cartesian spaces. However, the body does not operate on Cartesian principles, so this apparent cleanliness need not translate to transparent, unambiguous interactions.

On the other side, brain-computer interfaces (BCIs), typically based on the electro-encephalogram (EEG), purport to offer a window into a person's thoughts,  detached from bodily activity. Rigorous work in neuroscience and the application of neurophysiology to HCI \cite{2010} debunk the common myths of being able to ``interact with your thoughts and emotions''. Paradigms of interaction include tracking attention by evoked response and human-robot interaction, effectively exteriorising internal cerebral activity. Leaving aside the discrepancy between the expansiveness of rhetoric and the low interaction bandwidth of available technologies, BCI discourse can easily come to reinscribe a familiar Cartesian dualism between mind and body, only with the technology positioned on the other side of the divide. 

By contrast, EMG signals and interfaces confound any dualist understanding. Neuromuscular signals can result from both voluntary and involuntary activity - one could say `conscious' and `unconscious'. The activity that leads to volitional movements is inseparable from both cognitive intention and body mechanics. Temporally, EMG signals precede observable body movements, with the curious effect that interactive systems based on EMG exhibit \hl{a seeming} ``negative latency'' with respect to the user's familiar experience of body movement \cite{sanger2007bayesian} \hl{and lag typical of external sensing by inertial measurement or motion capture}.

\subsection{The EMG in HCI and Artistic Practice}

The use of muscle-based interfaces in HCI has been motivated by the need to interact with a nonphysical interface \cite{putnam_real-time_1993}. Applications have been proposed for users with disabilities \cite{barreto_practical_2000, putnam_use_1993}, in wearable contexts where devices are too small to embed traditional physical interfaces \cite{wheeler_gestures_2003}, for motionless subtle interactions \cite{costanza_intimate_2007}, or to allow interactions while busy with other tasks \cite{saponas_enabling_2009}. EMG in HCI is not limited to hand gestures and has been used on the face for speech recognition \cite{manabe_unvoiced_2003} and measuring frustration \cite{hazlett_measurement_2003}.

\hl{In the musical domain, EMG setups have been used across a wide range of styles and genres. For instance,} well-known performance artists Laurie Anderson and Pamela Z used the Bodysynth system in the 1990s to integrate EMG interaction into their multimedia stage performances \cite{kalvos__damian_cec_nodate}; Yoichi Nagashima used EMG to extend the performance of Japanese traditional music \cite{nagashima_bio-sensing_2003}; and 
Karolus et al. used EMG to augment the guitar \cite{karolusEMGuitarAssistingGuitar2018} and piano keyboard \cite{karolus:2020}.
More recently, the introduction of commercial devices like the Myo gesture control armband made EMG more accessible and led to a surge in interest in applications for artistic research \cite{nymoen_mumyo_2015}. Despite continued interest from the research community, \hl{the Myo device was discontinued in 2018 after only three years on the market, and its EMG-related patents were acquired shortly thereafter by Meta (Facebook's parent company) when they bought the neural interface startup CTRL-Labs.} There remain, however, several systems in the DIY electronics space providing basic EMG functionality \cite{da_silva_bitalino_2014}. \hl{We have used some of these systems, but, often out of necessity, have also conceived and built our own custom EMG hardware for the studies presented below, including the EAVI EMG board }\cite{tanakaeavi, tanaka_cmj2023} \hl{and the VoxEMG} \cite{Reed_VoxEMG_2024}. 

One question that arises from examining the sheer volume of \hl{EMG} implementations in HCI is: Why do EMG researchers and projects come and go? EMG is presented as an exciting, boundless opportunity to connect humans directly to their bodies. And yet, many researchers work with EMG for a little while and then move on --- What do they think they are getting into, and why do they quit, leaving their EMG research products and knowledge to fall by the wayside?

To be sure, EMG poses a wicked problem: working with EMG is challenging and the temptation to move to a less stochastic biosignal can be strong. But why? Although EMG is often framed by HCI researchers as a smaller, more subtle, sensitive input to carrying out classic interaction tasks, we argue that EMG is fundamentally different than the Cartesian basis for physical interaction, such as IMUs in tilt sensors or motion capture systems. Below, we retrace our own evolution in thinking from our initial attempt to use EMG in a classical human-interface control paradigm to one that accommodates its liveliness and `messiness'. \hl{Co-authors Tanaka, Fierro, and Reed have worked with EMG in the design of interactive musical interfaces. Tanaka has adopted the EMG in studying musical gesture for over ten years, focusing on the forearm and upper arm; Fierro has worked for three years with composers and acoustic instrumentalists creating EMG/EEG digital instruments; and Reed has designed EMG-based vocal performance wearables, both for other singers and for her own performance, for nearly five years (n.b., their respective projects are noted in the Acknowledgements section).} The three case studies presented here share a \hl{grounding in music}, but differ in electrical instrumentation, bodily configuration, engineering implementation, and \hl{conceptual and aesthetic basis}. In each, we identify how our design processes, coupled with the artistic deployment of the results (if indeed design and deployment are separate), have changed our views, understanding, and appreciation of EMG in interactive systems to find common themes in addressing its nature.


\section{Case Studies}

\subsection{Body as Instrument}
\begin{figure*}[ht]
        \centering
        \begin{subfigure}[b]{0.32\linewidth}
            \includegraphics[width=\linewidth]{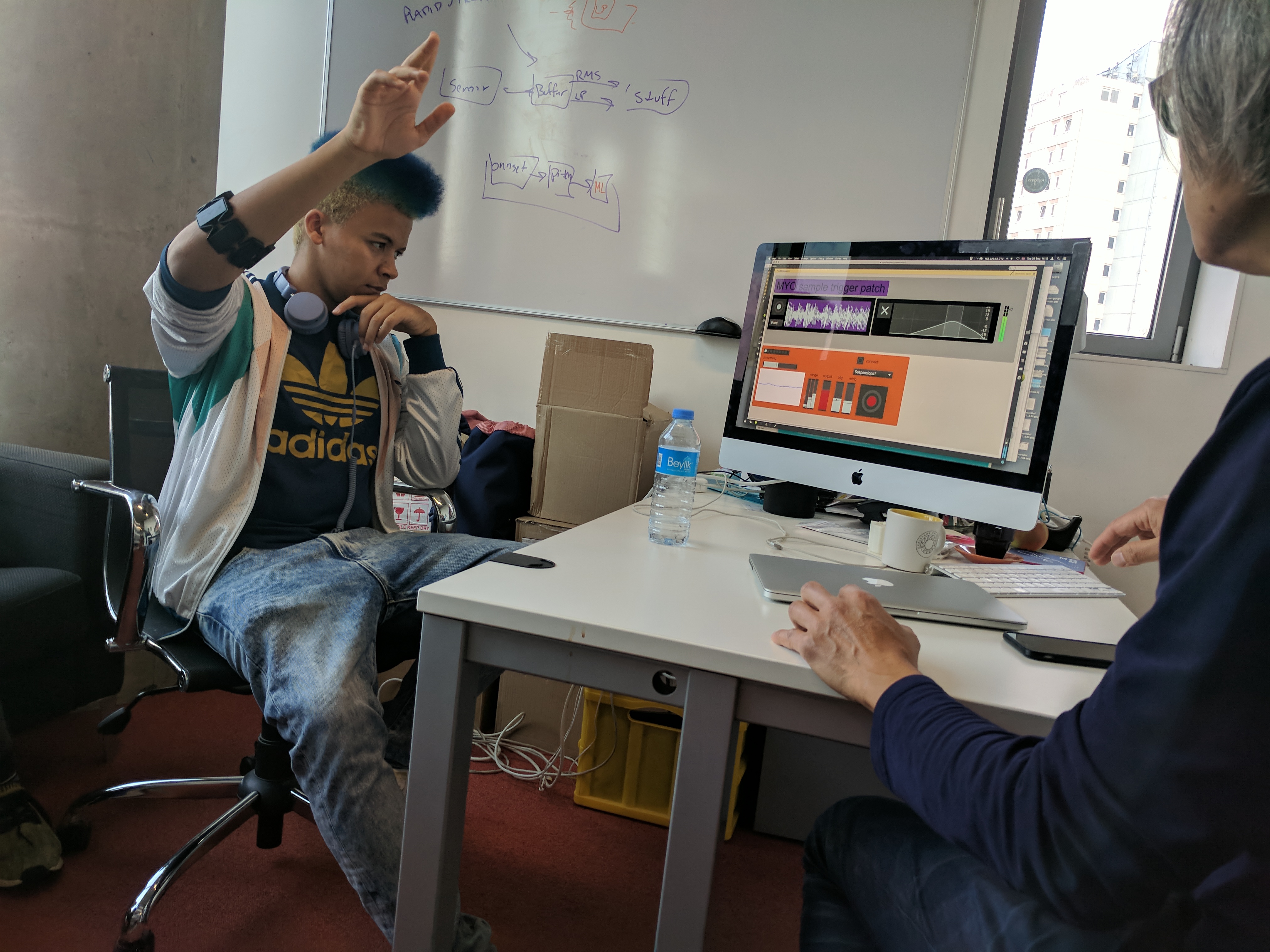}
            \caption{Musician from Blue Lab Beats triggering percussion sounds with EMG.}
            \label{fig:bluelabbeats}
        \end{subfigure}
        \hfill 
        \begin{subfigure}[b]{0.32\linewidth}
            \includegraphics[width=\linewidth]{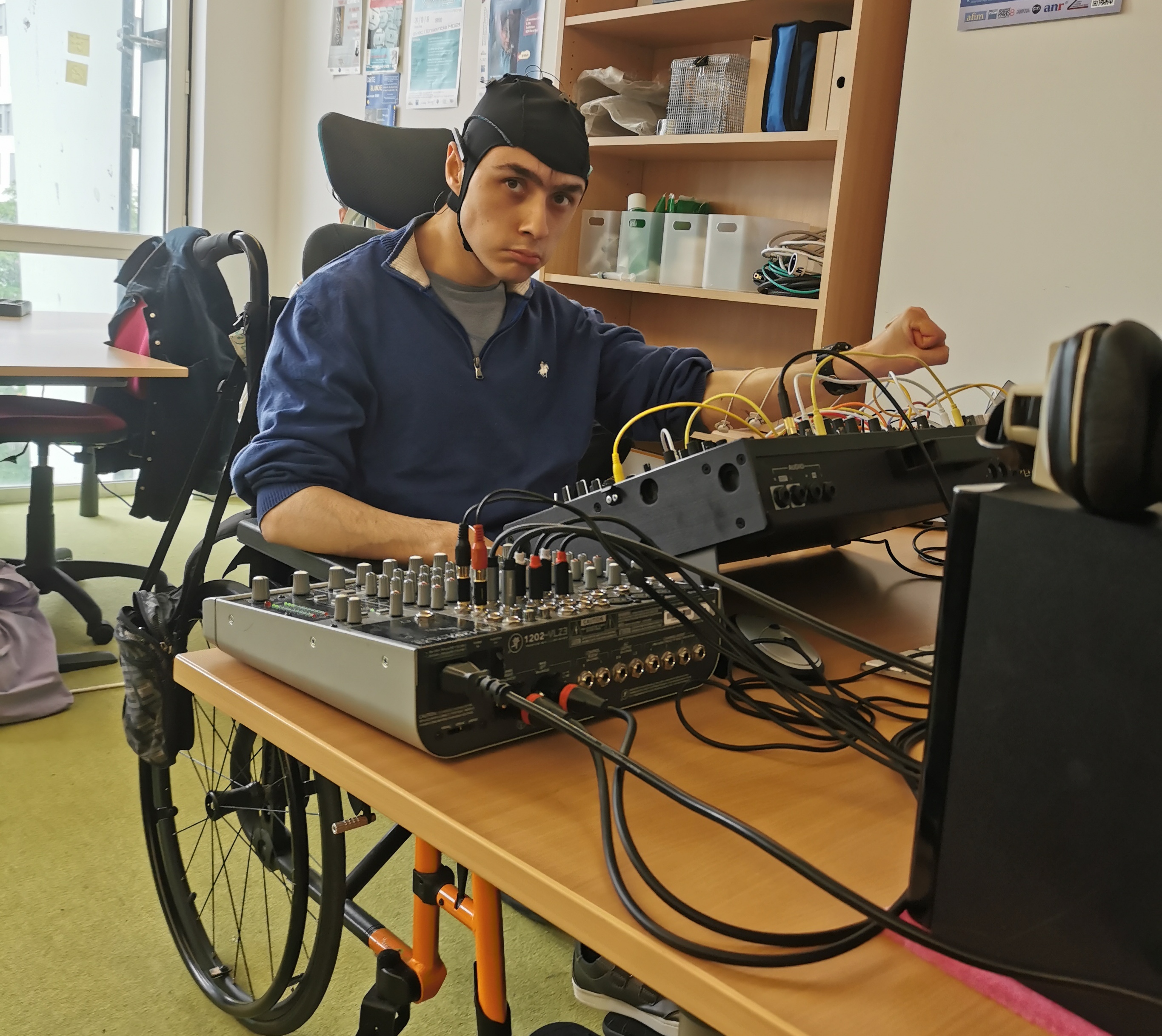}
            \caption{Autistic musician Cicanoise testing lab EMG system with synthesizer.}
            \label{fig:cicanoise}
        \end{subfigure}
        \hfill 
        \begin{subfigure}[b]{0.32\linewidth}
            \includegraphics[width=\linewidth]{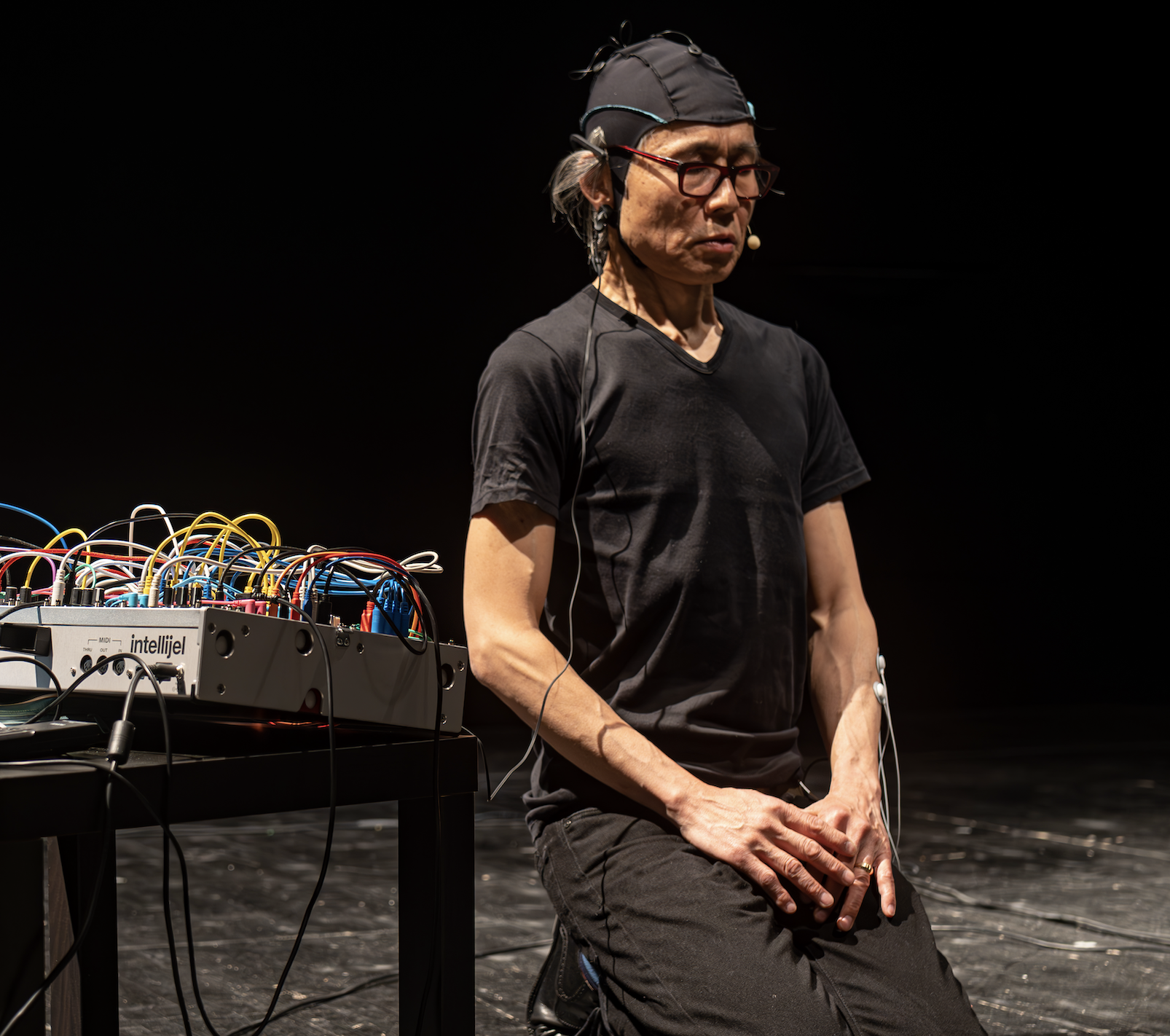}
            \caption{Tanaka in EMG/EEG modular synthesizer performance.}
            \label{fig:modular}
        \end{subfigure}
        \caption{Body as Instrument: From control to neurodiversity to modular.}
        \label{fig:Case_study_1}
    \end{figure*}

In this section we describe Tanaka's use of EMG on the forearm muscles to create gesture-based sonic interaction. In this case, he does not track muscular activity of existing musical practice associated with playing traditional instruments, but rather creates ``air instruments''. He explores four modes of interaction: 1) control features, 2) direct sonification, 3) machine learning, and 4) modular synthesis. These modes have evolved over time in response to his work with users, as well as to his following and adopting the continuous development of new interaction paradigms in embodied HCI. There is, in each instance, an auto-ethnographic component as he first tries a new interaction modality, then works with users of diverse musical styles and physical abilities (Figure \ref{fig:Case_study_1}).
 
\subsubsection{Control Features}
The first interaction mode is event based and draws upon a human-interface device (HID) paradigm exemplified in the computer mouse - pointing a cursor in a Cartesian plane, and selecting by clicking. This paradigm was prevalent in early digital music technology, exemplified in the MIDI protocol, where event types include Continuous Control (CC) mimicking cursor movement, and Note On/Off mimicking clicking events. Knobs and sliders illustrate CC while the piano keyboard provides the clearest metaphor for Note on/off events. Other musical instrument types, such as stringed instruments and woodwind instruments were adapted to this paradigm to produce `controller' devices emulating guitars and saxophones. A similar adaptation is necessary to adapt the messy, information-rich EMG signal to create tidy events and control messages.

The process of adapting the living EMG signal to one conforming to the HID paradigm involves smoothing and filtering, generalised as forms of signal treatment and processing called `feature extraction'. Features constitute meta-data information derived from the data-as-signal.\footnote{\hl{HCI research, including some of our own previous papers, sometimes uses the term `raw signal' to denote the signal captured from electrodes. The term should be seen as} a convenience to draw a contrast with feature extraction and further signal processing, without presupposing that this electrical signal is more neutral or closer to an underlying truth about the body than any other treatment.} A comprehensive review of EMG features is found in \cite{phinyomark2013emg}. Tanaka used various techniques of amplitude estimation \cite{sanger2007bayesian} to track the intensity of muscle contraction, from which he could derive a threshold-based trigger to generate an event, and filter the EMG signal to get a smooth contour that followed the exertion made by the user. By using two channels of EMG on two muscle groups, he could thus emulate a mouse to get X-Y movement and clicking, or create and ``air violin'' where continuous tension selected the pitch, and a note event articulated, or ``bowed'' the note. 

 This first mode of interaction followed electronic music traditions established by the Theremin to allow the musician to perform sound using their muscles to make free gestures in space. Tanaka's own research moved from emulating instruments (air violin, air piano, air guitar) to creating an embodied Theremin, to exploring entirely new sounds. \hl{These examples were shown publicly as technology demonstrations and also performed in professional musical concert situations} \cite{tanaka_cmj2023}. 

 The control feature paradigm was also tested on musicians producing music on computer-based systems as an alternative to commercially available interfaces (typified by keyboards, drum pads, and computer keyboard/mouse input). Tanaka created an EMG-based trigger/control system for finger drumming and sampler performance with the London-based jazz hiphop band, Blue Lab Beats. The band members have a virtuosic practice of rapid finger triggering of percussion and voice samples on MIDI pad interfaces, shaping the timbre of the sound with the other hand on sliders and knobs. They expressed a desire to free themselves from the device to create unencumbered arm gestures to play their samples. To realise this goal, a threshold-based trigger mechanism was created, taking care to use hysteresis to avoid unwanted re-triggering, while using parallel input to select different samples and apply continuous effects such as low-pass filtering. This style of music made demands on the latency and responsiveness of the EMG system. The system needed to trigger a sample at the precise moment the musician made a deft contraction in the muscle. More importantly, it needed to take place at the moment the musician felt right. This required some adjustment of the smoothing and trigger threshold. One user engaged in a productive learning process where, over the course of a one-hour studio session, he was able to more consistently and quickly re-perform one of his existing tracks. This created what Mackay describes as a \textit{co-adaptive} situation \cite{mackay2023creating} but also pointed out the limits of the control features interaction paradigm.

\subsubsection{Sonification}
If the control features paradigm sought to adapt the EMG to existing HID paradigms, the second interaction mode, sonification, provided a way to remove the potential loss of information that feature extraction imposed. The sonification (or, more precisely, audification) technique follows the paradigm of data visualisation, where data phenomena are made apparent by mediation and translation to other representational modes. 

 The EMG data is treated directly as an audio signal, and presented to the listener in brute form. Given the nature of the EMG as a stochastic pulse train, \hl{it is an aperiodic signal that does not exhibit fundamental qualities of musical tones, which typically have a periodic frequency and a series of harmonic overtones.} The encounter of listening directly to the EMG can be compared to listening to a modem or fax making a data connection on a phone line, or to seismic activity. \hl{Elsewhere, Tanaka details strategies for rendering the EMG signal as sound, and he describes a framework to treat the EMG in the audio signal processing chain} \cite{tanaka_cmj2023}. 

 With this interaction modality, all muscle activity – volitional and involuntary – are heard. Increased muscle exertion results in a greater number of myoelectric impulses to be generated by the body, producing a crackling sound that intensifies in amplitude and density. This has been adapted in concerts by Tanaka as part of a dis-intermediated, experimental body-sound performance setup. While it may be considered `music' in avant-garde circles, the unprocessed EMG data is not inherently musical in any conventional sense.

 Tanaka has identified several strategies for capturing the directness of the EMG data to produce harmonic musical sound. These techniques were composed in a compendium, much like a classical musical étude, and performed in public settings. These strategies included using the EMG data as a stimulus to excite resonant filters and physical models of acoustic instruments. The resonant filters were akin to making a clapping impulse in an acoustically rich tunnel. Similarly, with the physical models, the EMG becomes the equivalent of the plectrum plucking a vibrating string or the lips buzzing on the embouchure of a trumpet. However, instead of a single pluck or breath, the impulse train of the EMG created a continuous excitation of these harmonic systems that tracked the physical exertion of the performer.

Audiences who attended these public performances described `feeling inside' the performer's body, or feeling that by watching and hearing the performer, that their own body was doing the same thing. Some more scientifically-minded audience members evoked the language of embodied cognition in describing their experience of a mirror-neuron effect.

\subsubsection{Machine Learning}
If control features attempted to simplify the EMG data while sonification sought to lay bare its complexity, both techniques confront the difficulty of manually processing this information-rich signal. Recent developments in machine learning and AI technology enable such mediation to be carried out automatically by information analysis algorithms. In this paradigm, it is said that one does not need to understand the information carried in the data, nor need to carry out feature extraction, as this task is handed off to a black-box. The paradigm of interactive machine learning \cite{fiebrink_introduction_2018}, where the user engages with the training set in working with high-dimensional data, holds potential for productively accessing the complexity of the EMG signal. 

 Tanaka used the Rapidmix family of software tools that allow musicians to work with machine learning tasks such as classification and regression \cite{bernardo2017interactive}. The EMG signal was the input (or `target') to the system, with digital sound synthesis the output. Examples were recorded with input/output pairs associating discrete EMG input with sound output, and these were used to train a neural network (`training mode'). The resulting regression model was then run in real time (`testing mode') with live EMG input to create continuously interpolated sound output. EMG input in effect allowed the user to sculpt sound through gesture. Rather than engage with a complex sound synthesis engine one parameter at a time, gesture input allowed the user to simultaneously modify multiple parameters at once, in a way that was based on examples they provided in the training phase. \hl{The strategies used in segmenting EMG input to produce discrete training examples is described in } \cite{visi2021interactive}. 

 Tanaka was then interested to work in a high-dimensional information space where qualities of gestural input could be mapped to auditory meta-data by means of the neural network. This allowed him to take multiple channels of EMG data on different muscles and to effectively navigate corpora of sounds that had been classified according to their acoustic features. It became a kind of multi-dimensional joystick to explore sound and so-called ``timbre spaces'' through gestural input, manifesting a system similar to that described by David Wessel in the late-1970s \cite{62c6b59f-3b29-3804-a95c-db9a1e9c1036}. 

 These systems were evaluated in workshop settings described in \cite{zbyszynski2019gesture}. This enabled novice users with no prior experience of EMG interfaces or of digital sound synthesis to carry out sound design through embodied interaction. 

\subsubsection{Modular Synthesis}
The fourth interaction mode Tanaka explored was that of modular synthesis. Modular synthesis was originally a paradigm for analogue sound synthesis where musicians `patched' together sound producing and processing modules (oscillators, filters etc.) in physical networks by means of patch cord cables. This facilitated an infinite combination of modules, and resulted in patches where distinctions between signal (audio) and control (`control voltage' or CV) often intentionally broke down. The modular interaction paradigm has since been adopted in forms of node-based software programming where boxes connected by virtual cords replace textual programming in creative coding environments like Max, Touch Designer and Unity. Meanwhile modular synthesizers have seen a resurgence of late with the accessibility of electronic design and reliability of modern circuit components.

Tanaka worked with two companies - Rebel Technologies and McPherson's Bela \cite{mcpherson2017bela}, both active in the new modular synthesis scene - to create interfaces for modular systems. He used the OWL signal processing framework \cite{webster2014owl}, which had already been used in a series of other modular products, to allow the EMG signal to be used directly in a modular network. In this way, the user's body, it can be said, `became' a module in a modular system, and in this way was a source of data just as any oscillator (audio) or ribbon controller (CV). Once in the system, the EMG signals could be processed to be heard directly (as in the sonification case), or shape other sounds (as in the control features case).  Tanaka created several setups using the EMG following classical modular arrangements, including filtering the EMG data, and using the EMG as an input to a ring modulator.

To test the system's adaptability to contexts of diverse abilities and neurodiversity, Tanaka invited musician Robin Dussurget (aka Cicanoise) to try it in workshop and in concert. Dussurget is a wheelchair user and an autistic musician who performs with modular synthesizers. A neural condition means that his dexterity and motor control of the hands and fingers make turning the small knobs on the modular synth difficult. To address this diffculty, the EMG apparatus was placed on his arms and patched the system into the modular system, thereby allowing him to gain an expressive space in performance. He has performed publicly four times with the system, and has presented the system and what it affords him in workshop settings with musicology researchers, conservatory students, and autistic young people.

\subsection{EMG Signals on Electro-Acoustic and  Mixed Music}
Within the ambit of the BBDMI project \cite{tanaka_brain-body_2023}, Fierro and Tanaka have developed a modular \hl{software architecture} \cite{dimaggio_AM2023} to streamline the process of crafting musical instruments using the body's and brain's electrical activity \cite{dimaggio_AM2023}. Over the course of this project, they have conceived, designed, and tested an array of prototypes, each contributing to our understanding of the potential and limitations of such technology.
The focus of this \hl{section} is to delineate their experiences and insights garnered through the development and application of EMG-based musical instruments. During the design of these instruments, they have delved into two primary methods of interaction between the performers and the musical results:

\begin{itemize}
    \item Body to Sound Effects: This approach leverages EMG signals to manipulate various sound effects, arranged to create a specific digital musical instrument. Using EMG activity to manipulate sound parameters enables performers to sculpt and manipulate soundscapes in real-time through their movements and gestures.
    \item Body to Synthesised sounds: Here, the focus shifts towards utilising EMG signals to drive synthesisers, transforming gestures into synthesised sound.
\end{itemize}

The construction of these digital instruments drove Fierro and Tanaka to think about what type of interaction they wanted to achieve with these instruments. Part of the design process of the instruments involved the EMG signal interpretation and the interaction between these signals and the resulting sound or sound effect.

\subsubsection{Designing Interactions for EMG-Based Musical Instruments}
The acquisition of EMG signals for their application as inputs in sound generation encompasses a variety of interpretative approaches. Initially, interpretation begins at the moment of signal capture, necessitating an integrated view of both the body and the measuring apparatus as a unified system. By examining the body and device in concert, the goal is to explore the diverse potential of EMG signals for musical expression, acknowledging the spectrum of signal fidelity and nuances offered by different technologies. \hl{This integration helps performers grasp the instrument as a cohesive entity, enabling them to translate physical gestures into desired musical outcomes more effectively} \cite{cook_eduardo_2007}.

    \begin{figure*}[ht]
        \centering
        \begin{subfigure}[b]{0.32\linewidth}
            \includegraphics[width=\linewidth]{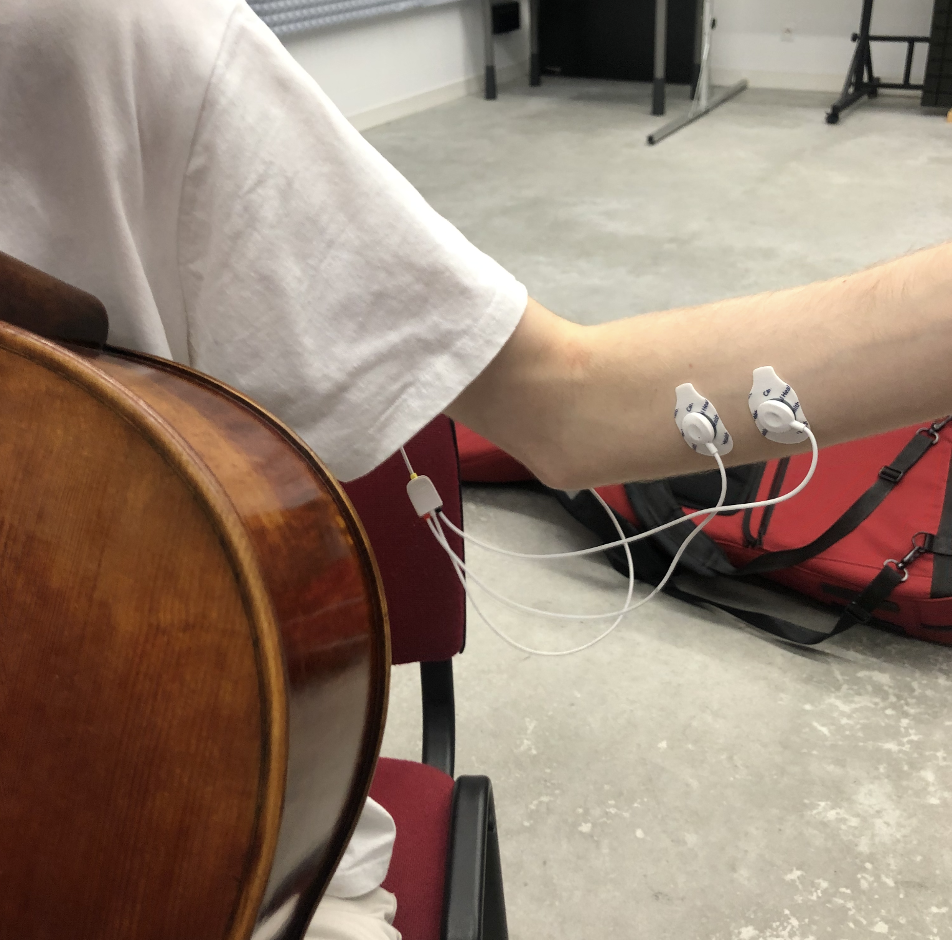}
            \caption{Cello player with electrodes on his forearm.}
            \label{fig:forearm}
        \end{subfigure}
        \hfill 
        \begin{subfigure}[b]{0.32\linewidth}
            \includegraphics[width=\linewidth]{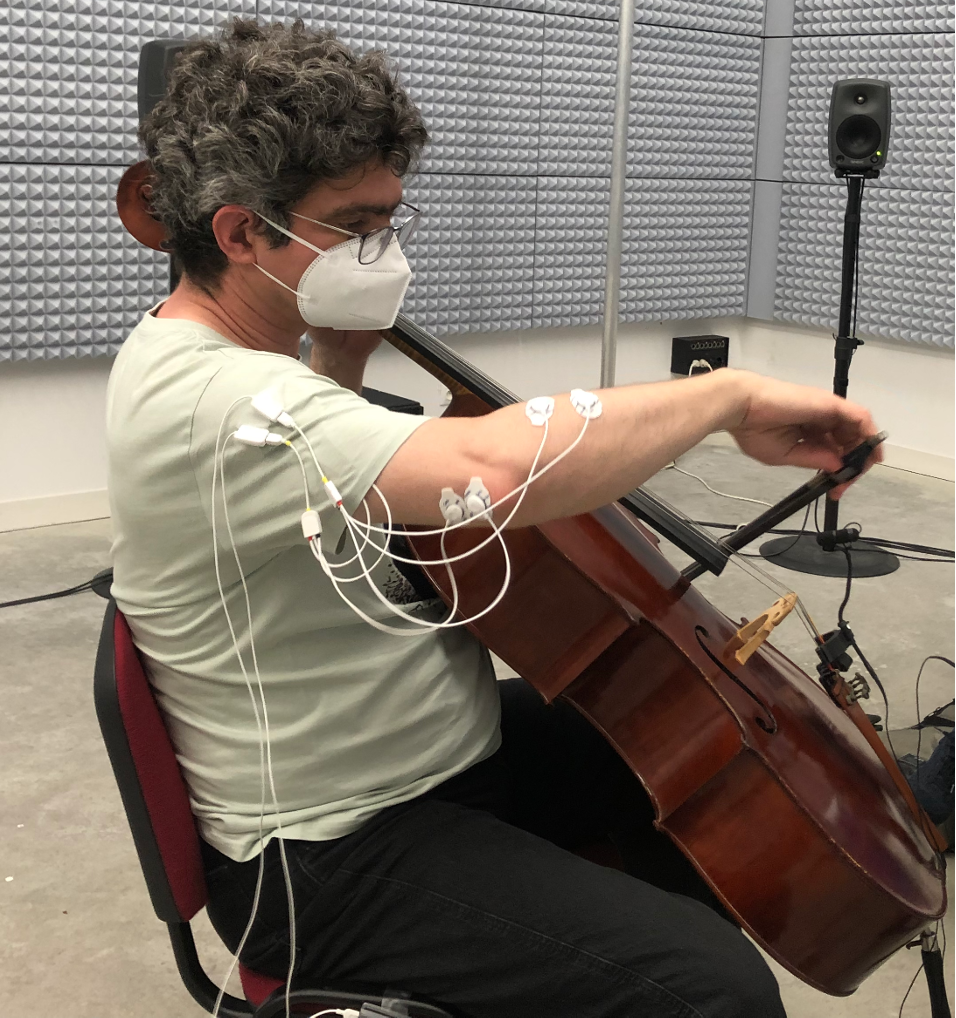}
            \caption{Cello player with electrodes on his forearm and elbow.}
            \label{fig:cello}
        \end{subfigure}
        \hfill 
        \begin{subfigure}[b]{0.32\linewidth}
            \includegraphics[width=\linewidth]{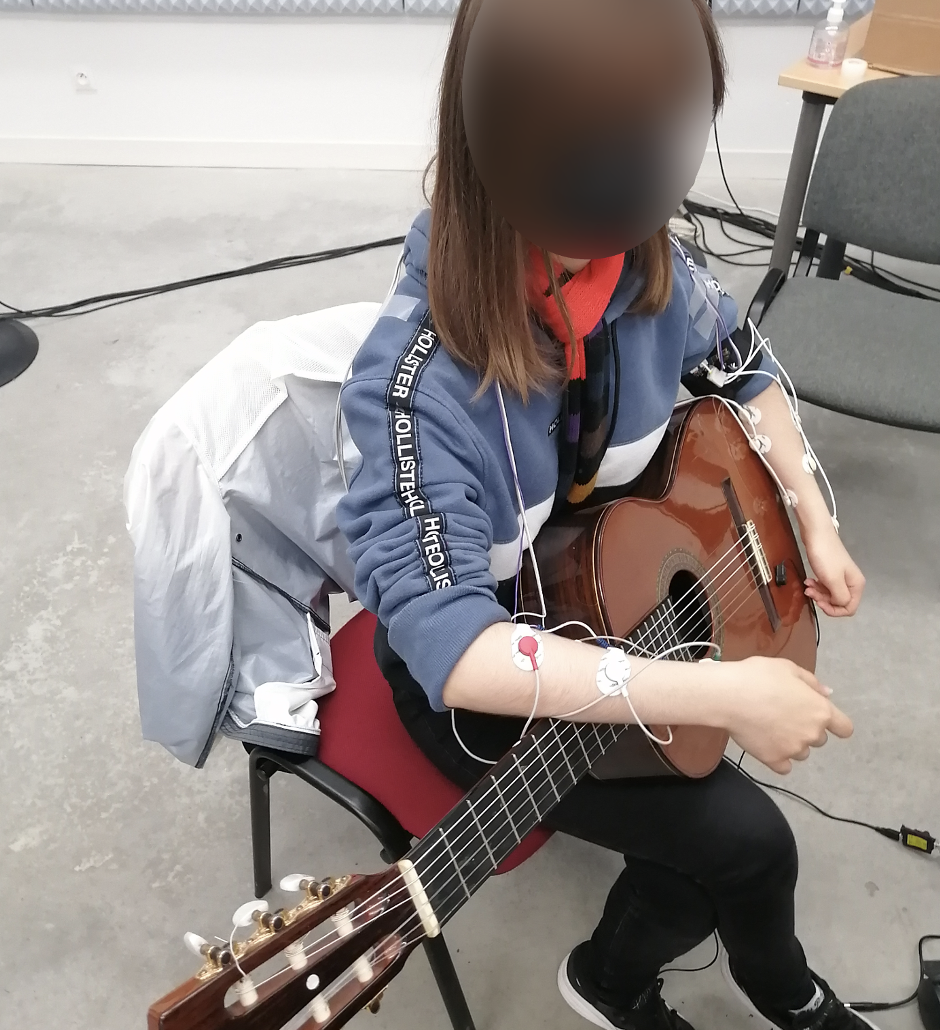}
            \caption{Guitarist with electrodes on both forearms.\\}
            \label{fig:guitarist}
        \end{subfigure}
        \caption{Musicians with EMG electrodes on gesture-involved muscles playing instruments.}
        \label{fig:active_muscles_electrodes_1}
    \end{figure*}
    \begin{figure}[htp]
        \centering
        \includegraphics[width=0.9\linewidth]{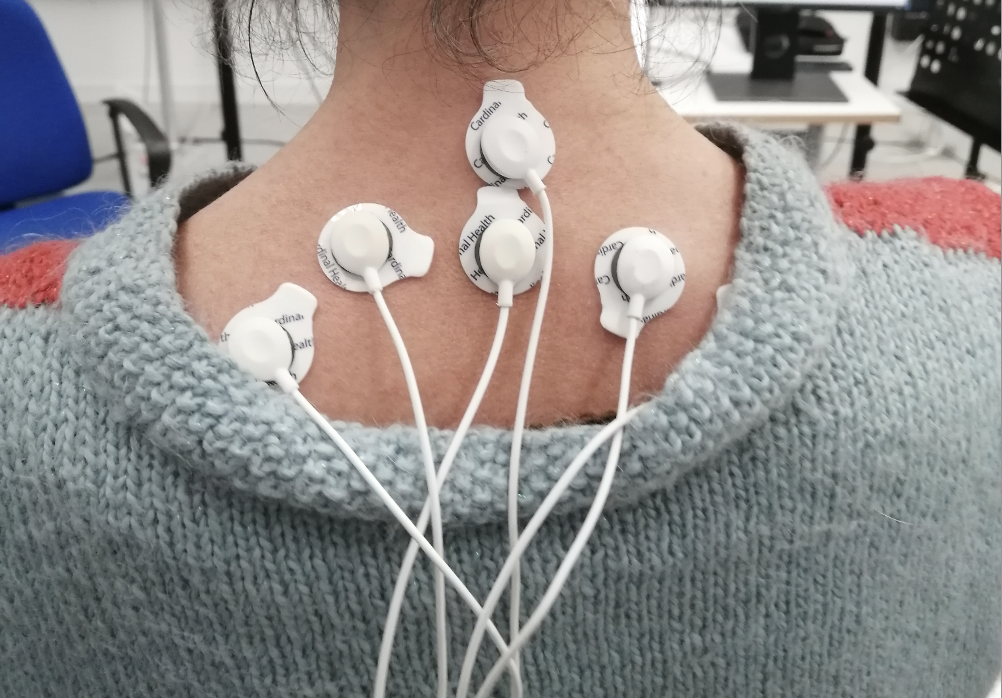}
        \caption{Guitarist playing her instrument with EMG electrodes on selective gesture-involved muscles.}
    \label{fig:trapezius}
\end{figure}

\subsubsection{Placement of Electrodes}
\begin{figure}[ht]
        \centering
        \begin{subfigure}[b]{0.49\linewidth}
            \includegraphics[width=\linewidth]{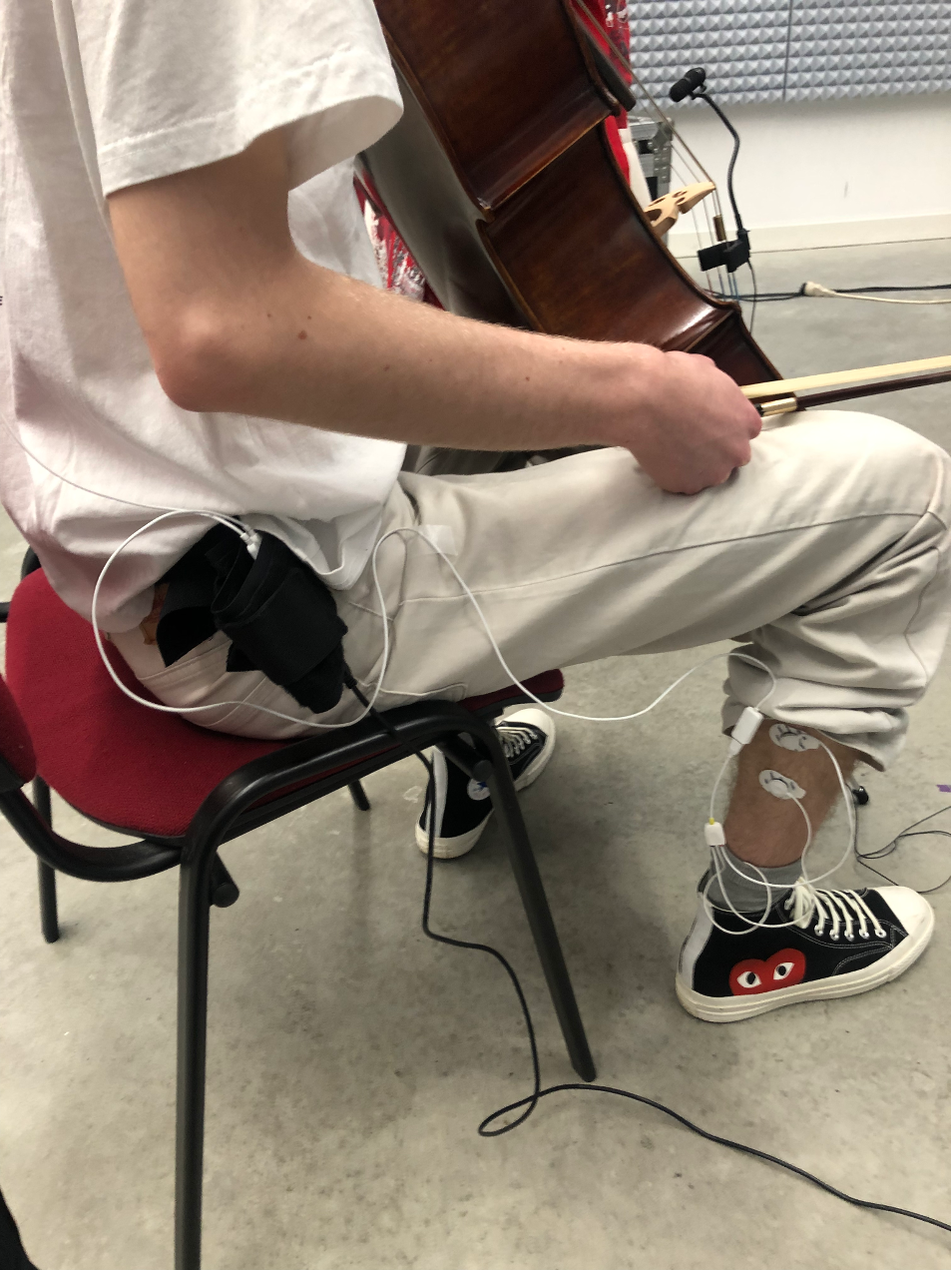}
            \caption{ }
            \label{fig:electrodes_leg}
        \end{subfigure}
        \begin{subfigure}[b]{0.49\linewidth}
            \includegraphics[width=\linewidth]{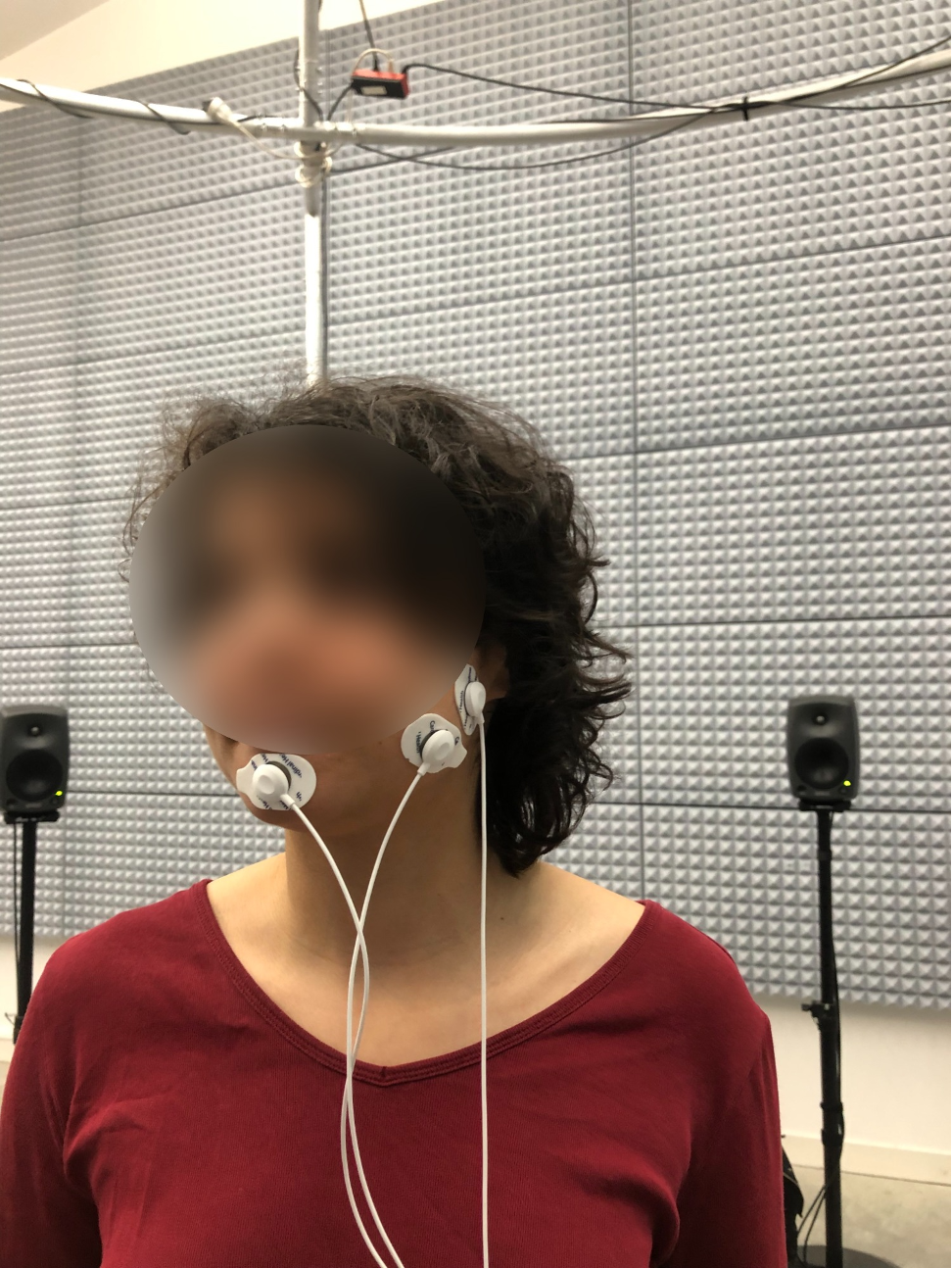}
            \caption{}
            \label{fig:electrodes_face}
        \end{subfigure}
        \caption{Musicians with EMG electrodes on gesture-involved muscles playing instruments.}
        \label{fig:active_muscles_electrodes_2}
    \end{figure}

Through their work, Fierro and Tanaka have identified two main reasons for using a limited number of electrodes in their designs. Firstly, the constraint stems from the limited availability of electrodes on their measuring equipment, which aligns with the standards of both commercial and open-source devices currently available \cite{Bongers2000PhysicalInterfaces}. Secondly, the challenge arises from the complexity and the significant learning curve required for performers to adeptly use EMG signals \cite{Donnarumma2011XthSense}. 
The necessity to work with a restricted number of channels requires inventive approaches to employ a minimal set of input signals. \hl{Given that electrodes serve as the sole source of input data, their strategic positioning is paramount.}
The process of electrode placement is a collaborative endeavour, involving input from both performers and the development team. Developers bring their expertise to inform performers about the effects of targeting different muscle groups. Often, performers find it challenging to anticipate their interaction with the system upon first use. Therefore, electrode placement \hl{is} typically revisited and refined after an initial testing phase to better accommodate the performer's \hl{existing instrumental practice}. At the outset, three distinct strategies for electrode positioning were explored with the performer, each significantly affecting their engagement with the system:

\begin{itemize}
    \item Placing electrodes over muscles engaged in performing instrumental gestures. This configuration implies that EMG signals are directly influenced by the actions required to play specific notes, as dictated by the score or the performer's immediate choice. This makes it challenging for musicians to manage playing precise notes while \hl{separately engaging the digital instrument}. Such a configuration allows for varied performance design approaches. In some instances, performers, when first introduced to the system without prior explanation, find their physical gestures and the resulting sound effects strictly determined by the music. However, through repeated use, performers gradually adjust their technique to enhance the quality of the sound effects produced, indicating a learning and adaptation process in their interaction with the digital instrument. Figures \ref{fig:forearm}, \ref{fig:cello} and \ref{fig:guitarist} depict two cellists and a guitarist with EMG electrodes on key gesture muscles.
    \item When electrodes are positioned on muscles that may, but do not always, engage in the act of \hl{playing their instrument}, this setup allows the performer to alter their physical movements to generate distinct EMG signals without \hl{impinging} their ability to play the intended notes. Performers make significant modifications in how they approach their \hl{acoustic} instruments when they are also managing a digital counterpart. Figure \ref{fig:trapezius} displays a guitarist with two electrode sets on her trapezius muscles, which may or may not be used for the instrumental gesture.
    \item By placing electrodes on muscles unengaged in the standard performance gestures, performers have the capability to intentionally invoke these muscles to generate EMG signals. This approach does not interfere with \hl{traditional instrumental practice}, thereby granting performers the freedom to manipulate the digital elements of the instrument independently, enriching their expressive range without compromising the integrity of their musical execution. Figures \ref{fig:electrodes_leg} and \ref{fig:electrodes_face} depict a cellist and a harpist with EMG electrodes on their legs and face, respectively, areas not involved in their instrumental actions.
\end{itemize}

In each setup Fierro and Tanaka have experimented with, participants have expressed experiencing the sensation of mastering an entirely new instrument. This arises from the digital component's requirement for either altering existing instrumental techniques, adjusting to new ways of engaging with the instrument, or learning to coordinate muscles previously not employed for musical purposes.
\hl{Continued exploration has underscored the necessity for EMG-specific musical notations. A straightforward approach involves allowing the performer's movements to directly influence the produced sound without establishing any explicit relationship between movement and sound. This approach fosters a sense of disconnect for the performer from the digital sound processing operations.} Alternatively, the composer might choose to give precise instructions to the performer on how to manipulate the digital instrument for achieving the intended effects of the piece. This could include detailing the exact EMG signal strength or motions needed. It's noteworthy, though, that requiring precise control over EMG signal output from those new to this technology can be daunting and might inadvertently distance the performer from the intended auditory experience.

The \hl{progression of proficiency of musicians engaging EMG-based instruments} is markedly influenced by the device's intricacy. When the connection between muscle activity and sonic output is straightforward, performers typically grasp the concept with relative ease. However, the multifaceted nature of the instruments introduces a complex array of interactive possibilities, elevating the challenge for musicians to become adept, thereby steepening the learning curve. In response to this, Fierro and Tanaka have devised several pedagogical techniques proven effective in our investigations. Initially, they present musicians with a visual depiction of the EMG \hl{data-as-signal}, encouraging experimentation with potentially performative gestures. Specifically, for instrumentalists incorporating EMG to modify sound, they suggest observing the signal feedback while engaging in varied playing styles. Newcomers often express astonishment at the dynamic between physical movement and EMG output—a revelation that substantially differs from preconceived notions. This preliminary phase of exploration significantly enhances awareness of how different gestures influence EMG intensity, aiding in their mastery of the instrument \cite{RefsumJensenius2017}.
\hl{The user cases demonstrate that the user's ability to learn how to interact with the system improves when they receive a step-by-step breakdown of their interactions. By enabling performers to isolate and experiment with individual electrodes and sound effects, they gain a deeper insight into the connection between physical movements and sonic outcomes.} Through practice, performers naturally link their gestures to the sounds produced, circumventing the need to consciously analyse the muscle-generated signals. As more \hl{complex} interactions are gradually introduced, connecting initial gestures to a broader range of sound effects, performers gain a holistic understanding of the instrument. This comprehensive familiarity allows them to develop a unique approach to playing the instrument, merging the physical and digital components into a cohesive whole.

\subsubsection{Designing the Interaction between Signals and Sound}
\begin{figure}[htp]
    \centering
    \includegraphics[width=\linewidth]{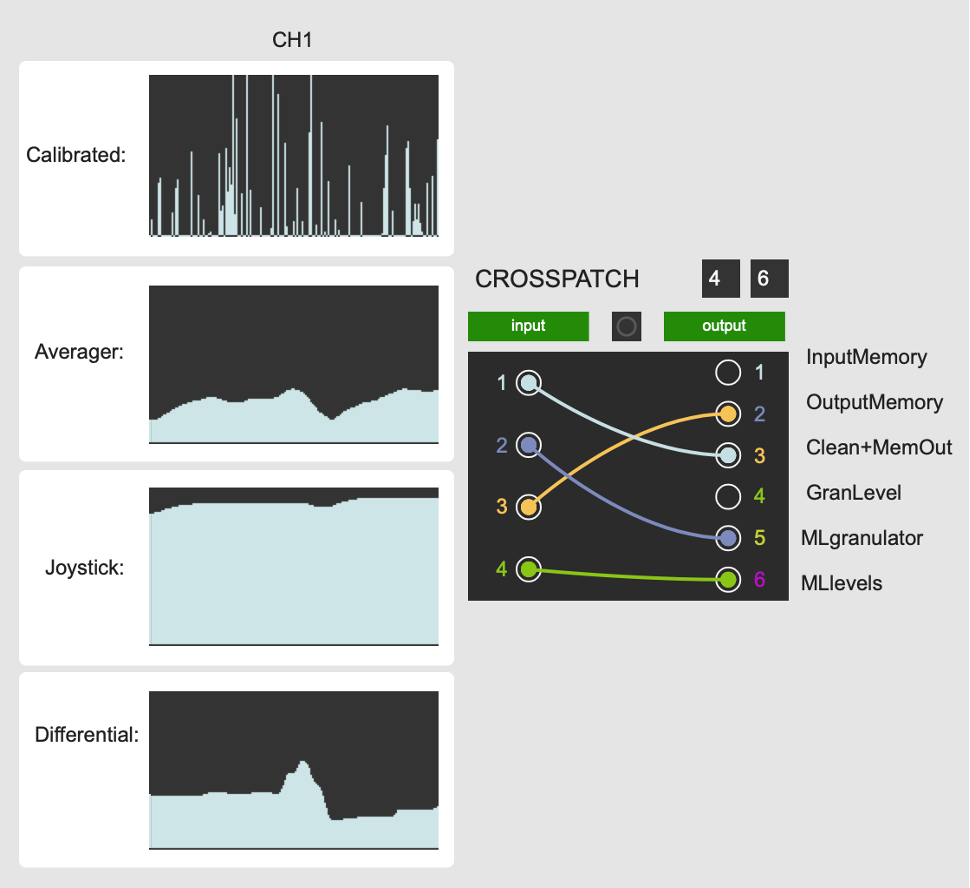}
    \caption{Various transfer functions from one EMG electrode to sound effects mapping.}
    \label{fig:transfer_funtions}
\end{figure}

As discussed in Section \ref{sec:emghistory}, sEMG signals are inherently stochastic, making it impossible to replicate an identical signal on two separate occasions. The characteristics of these signals are an amalgamation of spikes influenced by various factors, including the nature of the gesture performed and even muscle fatigue. Integrating this unpredictable nature into the design process presents certain challenges, yet it also introduces an element of ``messiness'' and natural variability that is hard to achieve with other data sources, such as external sensors not directly connected to the body. In response, Fierro and Tanaka have designed instruments to capitalise on this aspect, offering users the opportunity to work with the \hl{unprocessed signal} or employ more complex transfer functions to harness this variability.
A notable challenge in utilising EMG data for performance is muscle fatigue over extended periods. When performers aim to generate strong EMG signals, considerable effort is required, even if the input signals are normalised at the outset of the interaction. To address this issue, Fierro and Tanaka have developed a modular system that enables users to dynamically choose among various transfer functions. These functions are designed to allow performers to achieve any desired output level with minimal physical exertion. The development of these transfer functions leverages an understanding of human gesture's temporal dynamics on both broad and detailed scales, facilitating a more sustainable and user-friendly interaction with the system. 
The ``joystick'' and ``differential'' modules exemplify the system's adaptability to EMG signal input. The joystick module functions akin to a virtual joystick, where the strength of the EMG signal dictates the output: exceeding a certain threshold increases the output, while maintaining the signal at a consistent level keeps the output steady, and reducing muscle tension lowers the output. This design not only facilitates achieving various output levels with less muscular effort but also enables users to sustain a specific output level for extended periods. On the other hand, the differential module responds to short-term fluctuations in the input signal by analysing its derivative, thus altering its output based on rapid muscle contractions (to increase output) or swift relaxations (to decrease output), offering a nuanced control mechanism based on the immediacy of muscular movements.
Figure \ref{fig:transfer_funtions} illustrates the system's modular design, enabling users to choose from a variety of transfer functions for system interaction and assign them to specific sound parameters. Through this setup, each electrode is treated as a source of multiple signals, rather than just a single output. As depicted in Figure \ref{fig:transfer_funtions}, a singular input level can generate several distinct outputs at various intensities, including an option that closely mirrors the \hl{unprocessed signal}. While this framework permits the creation of intricate instruments with numerous adjustable features and minimal input electrodes, empirical observations have highlighted a significant learning curve. Performers typically require extensive practice to adeptly navigate the instrument and fully exploit its capabilities.

\subsection{Vocal EMG}
Finally, we describe vocal EMG through Reed's autoethnographic account of designing for the musculature of her larynx and further vocal EMG work with other vocalists. This work delves into three aspects of vocal EMG: 1) Representing vocal laryngeal movement, 2) Collaboration with the noisiness of the vocal body, and 3) (Mis)alignment between EMG sonification and singers' expectation and understanding of their voices. 

\subsubsection{Representing the Voice (or so we thought...)}
Reed's work with vocal EMG began with a preconceived notion that the body is controlled by the vocalist to enact precise technical manoeuvres. The ``success'' of this execution is assessed by the vocalist based on the difference between expected and actual sonic feedback. In her initial EMG implementations and experimentation, Reed's inclination was therefore to explore vocalists' auditory imagery and response to auditory feedback. She believed this to be the most critical mapping --- of sound to action --- in use by vocalists; this belief was formed in Reed's background in classical Western vocal pedagogy and practice, which consequently forms her understanding of her own voice, \hl{and inherently her design for the voice.}

The voice is mainly observed through auditory means (e.g., analysed audio recordings as an input for vocal control and consequently why auditory feedback is a common mechanism when designing for vocalists).
However, audio-based methods, like the other exterior (objective) methods mentioned in Section \ref{sec:sensormodality}, are somewhat indirect, telling us more about the results of action, rather than the action itself. In vocal performance, different physical movements can manifest in similar audio signals.
In contrast, EMG observes the source of action, in what Tanaka describes as a ``precursor'' to observable movement (i.e., audible action of the vocal body) \cite{tanaka:2015}. 
The movement of the laryngeal muscles is visually unobservable and, although part of the vocal technique, is not overtly felt, rather being understood through the vocalist's tacit knowledge of their body. Therefore, EMG provides a way to peer inward at these typically unobserved, delicate, and precise movements.

Reed developed a custom PCB, the VoxEMG \cite{Reed_VoxEMG_2024, Reed_NIME20_VocalsEMG}, to measure action potentials across the small extrinsic muscles of the larynx. Differential amplification is used to acquire the neuroelectrical voltages and for \hl{common mode rejection} \cite{Reed_NIME20_VocalsEMG}.
The goal was to use EMG signals to investigate vocalists' imagery and intention, and potentially build classifiers for subvocal musical technique. \hl{The inspiration came from} AlterEgo, another vocal EMG platform developed to classify subvocal speech \cite{kapur2018}. \hl{Another case of disappearing media, AlterEgo had garnered attention as one of Time Magazine's Best Inventions of 2020, allowing users ``to communicate with your computer without touching a keyboard or opening your mouth'' }\cite{TimeMag_AlterEgo}. \hl{Reed's initial aim was to similarly classify musical vocal muscle movements and use them control specific sonification processes;} e.g., performing a particular vocal exercise would achieve a set sonic result, lending to the identification of an action or intention in an audible way.

Reed's initial implementation of the VoxEMG was therefore via auditory feedback. The EMG signal was mapped to sonifications, such as filtering ambient noise, thereby using sound to externalise vocal movements. Through development, a shift occurred in the relationship between Reed and her voice. \hl{She noted disparities between her assumptions of her movements and the reality of the EMG sonification.} What became more interesting, rather than classification or identification, was how unidentifiable or seemingly random movements were, at least at first. 

In later reflection, Reed noted her frustration was a result of a \hl{mismatch between her understanding of her body and the functionalities of EMG}: 
Certain vocal actions, such as singing different pitches, were expected to produce noticeable changes in the sonification but did not. 
Unexpected feedback required unpicking vocal technique and took time; \hl{for instance, Reed had a particular breakthrough when realising that activations she observed from her omohyoid muscles were part of her inhalations at points prior to vocalisation}; previously, she had not connected this action to her breathing \cite{Reed_TEI21_sEMGPerformance}. 
By giving attention to an isolated muscle, the low-level movements of the body became apparent and even startling. Exploring her movements through this novel feedback, Reed's behaviour and relationship with her voice became entangled with the EMG signals:

\subsubsection{Learning to Collaborate with the Unexpected Noise}

Reed brought her background in romantic and baroque opera into this interaction. Typically, classical Western vocal styles revolve around notions of control over the body, in contrast to more modern techniques that often favour somatic approaches to felt experience in the body. 
\hl{Adding the EMG sonification into her experience, Reed's attitude towards these self-imposed technical constraints loosened.} Her vocal movement began to incorporate bending and twisting of the neck, rapid, shallow breathing, and breathier or raspier sound qualities from loosening control over the larynx. Through these actions, the EMG sonification takes on a life of its own and trends towards the experimental and playful. 
The EMG sonification provided a backdrop against which Reed's unaltered voice could still be heard but allowed her to take advantage of pushing and playing with her vocal technique. Emphasis was placed on collaboration with the vocal body as a source of change and uncertainty, as well as appreciation and engagement with its messiness.


Reed used an autoethnographic approach to study her experience of this design process, outlining how this externalised movement impacted her perception of her voice \cite{Reed_TEI21_sEMGPerformance}. Normally, 
no focus is given to an individual muscle and the movements are not felt except through interoceptive sense.
By examining a single muscle in isolation, the monolithic experience of the larynx was disrupted. 




This led to a sensation of the body as a collaborator. Rather than an element of control, the EMG sonification revealed how the self (directives and intentions of the singer) manifest in unconscious or unnoticed ways in the body. This provided sensations of the body as an uncontrollable ``other''. Noisiness is further compounded by the messiness of EMG signals taken from the throat: crosstalk from major arteries and auxillary gestures for posture during vocal performance can also be observed. The muscles support more than one functionality, meaning one-to-one mappings between EMG signals and sonifications can be unclear. Likewise, Reed's training may contribute to overall less muscular tension, making some activity undetectable. This noisiness and ambiguity led to a practice of collaboration and improvisation with the voice, wherein the body's presence added to the performance in unexpected ways.

\begin{figure}[t]
    \centering
    \includegraphics[width=\linewidth]{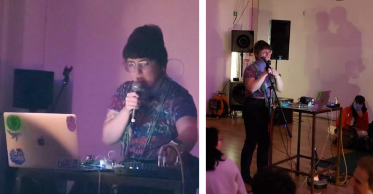}
    \caption{Reed performing with the \textit{Singing Knit}, an eTextile wearable using VoxEMG boards for laryngeal muscle interaction \cite{Reed_SingingKnit}, at IKLECTIK, London in April 2022.}
    \label{fig:c-iklectik}
\end{figure}

We reflect on how this has changed Reed's perception of her body and behaviour when returning to her typical, classical vocal context. Rather than readopting the view of the voice as a tool to be mastered, this feeling of collaboration has become entangled in her current vocal practice. \hl{In fact, this exploration of noise and unexpected mess is now a point of focus in Reed's artistic practice (Figure }\ref{fig:c-iklectik}). Having made the body strange \cite{loke-movingmakingstrange}, emphasises the power and contribution of the voice. Rather than trying to reign it in or command it, Reed's perspective is now of her voice as a partner, limiting control as a positive factor to be approached with empathy and care. This outcome might lead one to evaluate EMG as a bad tool for measuring technique; in interacting with it, technique changes --- the phenomena of vocal practice cease to exist and can only be observed in the way they used to be.

\begin{figure}[t]
    \centering
    \includegraphics[width=\linewidth]{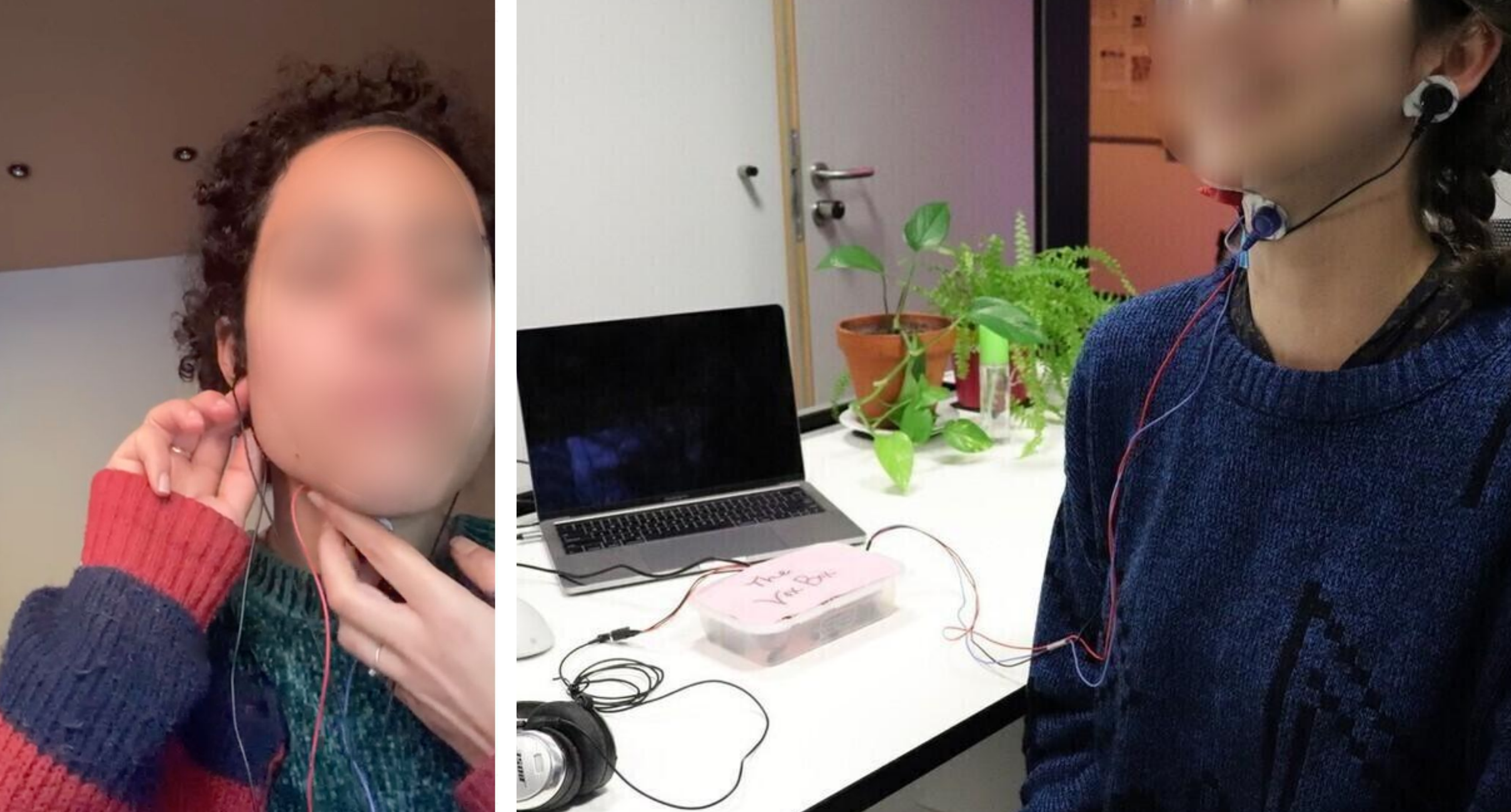}
    \caption{Vocalists using the VoxBox experiment setup for vocal EMG interaction \cite{Reed_TEI23_BodyAsSound} to set up their electrode connections on suprahyoid laryngeal muscles.}
    \label{fig:c-voxbox}
\end{figure}

\begin{figure*}
    \centering
    \includegraphics[width=\linewidth]{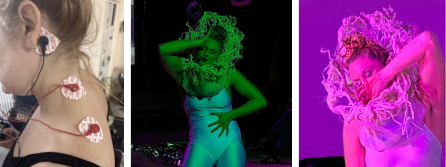}
    \caption{Performing Artist Lou Barnell's vocal EMG setup (left) and the sculpted collar for \textit{To Fly Out in the Heat of Day}, as performed at Theatre Deli, Sheffield in February 2024 (centre, photo by Emma Houston; right, photo by Anthony Bennett).}
    \label{fig:lou-collab}
\end{figure*}

\subsubsection{(Mis)aligning Vocalist Expectations}
The body's enaction of a particular practice may not necessarily align with human perception of that action. \hl{Reed's original auditory feedback representations, although predominant in her pedagogical domain, did not manifest as expected. Movements exist where they are not noticed and the body works in the background; not all vocal activity and muscular movement is heard and many movements are chained together to produce what is}. Similarly, personal perceptual mechanics and action-sound mappings are not necessarily \textit{factual} representations, although they constitute the individual's reality. \hl{Reed's exploration continued to examine others' experience with the sonification}\cite{Reed_TEI23_BodyAsSound}\hl{, as well as in the design of vocal EMG wearables around other devices and mappings.}

\paragraph{Individual Enculturation}
Reed used the same noise-filtered EMG sonification in a long-term study with two vocalists from different pedagogical backgrounds: Hindustani classical and jazz improvisation. The VoxBox toolkit was given to the vocalists (Figure \ref{fig:c-voxbox}), who worked the EMG into their practice \hl{over two months} \cite{Reed_TEI23_BodyAsSound}. Using semi-structured and micro-phenomenological interviews, Reed worked with the vocalists to \hl{underpin tensions in their experience with the sonification}. They experienced some of the same surprises as Reed, for instance the attention drawn to the non-vocalised breath. However, the other vocalists, bringing different vocal practice into the mix, \hl{faced different tensions. For the jazz vocalist, sensations of ``otherness'' from the sonification at times led to playful experimentation, and at others doubts about her vocal practice.} For the Hindustani singer, in a vocal culture demanding precise awareness of pitching, the inclusion of a non-pitched sonification was overwhelmingly disruptive. Further, the auditory feedback misaligned with the vocalist's very visual and somatic conceptualisations of her body and sound. We see how Reed's assumptions of vocal performance and entanglement with her own experimentation of EMG were inadvertently imposed on other singers with different bodily perceptions and musical cultures.

Perhaps the most interesting observation was the vocalists' expectation of the system to evaluate vocal technique as being ``good'' or ``bad''. 
Where Reed had hoped for and encouraged freeform exploration, \hl{there was an underlying desire for the sensor and system to evaluate something about the self or align with individual values in artistic practice} \cite{Reed_TEI23_BodyAsSound}.
This demonstrates some preexisting, encultured aspects of vocal evaluation. Like Reed, the other vocalists brought particular values to the interaction, such as the Hindustani singer's valuation of musical pitch. These are the expectations of the voice and therefore the technological system: ``am I using it correctly?'' EMG again seems ill suited towards evaluation; the noisiness of the body and what the EMG actually provides is misaligned from our conceptualisation of action-output mappings.

\paragraph{Devices and Mappings} \hl{Disparities are further influenced by the devices that coproduce the signal and mapping. In a recent collaboration between Reed, performing artist-vocalist Lou Barnell, and artist-researcher Robin Foster, the VoxEMG system was reimplemented within a sculpted collar to be worn by Barnell (Figure }\ref{fig:lou-collab}). \hl{The project, \textit{To Fly Out in the Heat of Day}, explores sculpting the physical and auditory aspects of Barnell's vocal body} \cite{Barnell_TFO}. \hl{Choreographed muscle flexes were intended to trigger parameters of a granulator and loop function in Max/MSP RNBO to manipulate a live vocal signal. The software patch was based on previous prototypes that used the Myoware (Advancer Technologies, NC) board for muscle sensing. Myoware's on-board processing scales EMG signals between 0 and +1V; designed around the Myoware, the expectations of the muscle became associated with the binary of this peak detection --- on or off, flexed or not. Conversely, the representation provided by the VoxEMG as a stochastic time series signal proved difficult to filter into thresholds to detect particular single muscle flexes. Deriving inspiration from these challenges, the project has now branched into two perspectives: one, using another physical prototype with Myoware sensors to represent movement in this functional binary, and two, reconfiguring the patch and sonification around the subtle, unexpected parts of the VoxEMG representation. These dual perspectives in body and EMG-focused design highlight how data measurements and processing constitute specific ideals and human perception of the body, as well as our values in performance and gesture }\cite{Barnell_TFO}.
\vspace{1em}

These vocal explorations demonstrate how EMG is entangled with perceptions and even misunderstandings of the measurement. Sonifications were based on Reed's notion that the vocal musculature could be mapped in a clear way from intention to activation to sound. The reality of the EMG is more complex and multidimensional; this notion, although cognitively relevant to the perception it was designed within, is reductionist in its own way. The signal does not exist in the way it is perceived. The VoxEMG and its implementations were not designed around the voice objectively as a measurable entity (such a task is not possible, given the encultured perception of the voice \cite{eidsheim2017maria}), but around Reed's vocal practice. To further complicate things, once interacting with the EMG system, the vocal practice itself changes \cite{homewood:bodies}. The EMG representation becomes entangled with subsequent perceptual mappings; as soon as it is introduced to measure the vocal phenomena, the phenomena cease to exist. In this way, EMG is not a very useful tool for examining vocal practice, but indeed for upending it \cite{loke-movingmakingstrange}. We can then examine what it \textit{used} to be, and how the \hl{representation} continually entangles itself into the voice.



\section{Discussion}\label{sec:discussion}

Our varied experiences in designing and performing with EMG systems have highlighted a number of characteristics of EMG as a material-discursive practice, sometimes with considerable divergence from the way the technology is typically conceptualised in HCI. These three case studies differ in their instrumentation, bodily configurations, signal processing and musical practices. Yet, all cases share certain common patterns. We here discuss \hl{theoretical considerations }\cite{wobbrock2012seven}\hl{ about EMG and data representationalism with suggestions for future designers and researchers wishing to incorporate EMG into their work.} 

\subsection{Tidiness and Messiness} 
A readily apparent feature of all three case studies is the messiness of the EMG signals. As discussed in Section \ref{sec:background}, this messiness could be ascribed to many sources, including the stochastic nature of neuromuscular signals, the limitations of sensing on the skin surface, and variability in how and where the electrodes are deployed by different individuals. The actuality of messiness stands in marked contrast to an idealisation of EMG as a window into interior bodily intentions. A classical engineering response might be to model the divergence from ideals as noise, perhaps reducible through more sophisticated instrumentation or signal processing, but at the very least characterisable with stochastic models. To an extent, some of our work deriving features from the electrical signals (e.g., envelopes) fit into this paradigm.

\hl{As an alternative approach}, following Sanches et al. \cite{sanches2022diffraction}, \hl{EMG provides an opportunity to move} away from abstraction and normativity toward working with \textit{this} data in all its specific messiness. 
\hl{Karolus et al. recently have engaged with this challenge for playful and ``body-insightful'' approaches to EMG and movement creativity focused on impreciseness }\cite{karolusImpreciseFunPlayful2022}. Examining trends in this work over time reveals the adoption of an EMG practice rooted in a \textit{playful} exploration. 
Leaving aside EMG as an (imperfect) signifier of abstract intention permits a closer inspection of the signal for what it \textit{is}, how it is experienced by the subject, and how it might differ between bodies and contexts of use. 
This is most apparent in our case studies where ``raw'' electrical signals are sonified with minimal post-processing; though again, the sense of rawness here comes not from a closer imagined proximity to the body, but from the simple refusal to infer conceptual meaning from the signal through digital feature extraction. Thus, while raw data may be a fiction \cite{gitelman2013raw} -- even the most straightforward sonification is not aesthetically neutral -- we find that leaving the meaning of the EMG signal to the auditory interpretation of a performer or listener can support productive forms of ambiguity \cite{gaver2003ambiguity}.

This is not to say that trivial sonifications are the only way to live with the messiness and ambiguity of EMG signals. As a material-discursive practice, the discourse surrounding EMG matters; intentions and narratives matter, particularly around why features are extracted and what those features notionally signify. \hl{Working with EMG, as well as other signals that purport to tell us something about the body, requires awareness and acknowledgement of how the data is ``tidied'' or not and why. For those who desire to work with EMG, the intentions must be made clear, especially to avoid misconceptualisations being carried over to an end user.} To extract a set of features for the purpose of detecting a set of well-defined gestures, which supposedly generalise across different bodies, is a different type of practice to extracting those same features for their intrinsic and varied qualities. However, the latter approach tends to cut against utilitarian, solutionist approaches commonly found in product design and certain corners of engineering research, instead being found (as here) in artistic contexts. \hl{As a rebuke to initial intrigue and idealisation as a view into the body, future design should acknowledge EMG's messiness and the process used to negotiate its ambiguity.}

\subsection{Sensor or Actuator? The EMG Beyond Representationalism}\label{sec:representationalism}

On first glance it seems obvious that EMG interfaces, in contrast to electrical muscle stimulation \cite{faltaous2022perception}, are sensors rather than actuators. However, an agential realist like Barad would argue that the observed is inseparable from the observer -- the act of measurement changes what is being measured \cite{barad2003posthumanist}. Even without adopting the full set of new materialist and quantum mechanical commitments of agential realism, our experience shows that EMG ``sensing'' displays many characteristics of an actuator. This extends, in some ways Lopes's work \cite{Lopes2015} on EMG stimulation as a source of proprioception. In our case studies, rendering neuromuscular signals audible invariably changed the behaviour of the person wearing the apparatus. 

Our case studies of instrumental and vocal practice showed how the performer's relationship to their own body changed rapidly once EMG signals were rendered audible in any form. This yielded different performance techniques, which in some cases persisted in contexts without EMG. 
In what might be described as a kind of \textit{experiential control} \cite{tuuri2017controls}, the sensing apparatus comes to alter the body that it senses. The measurement thus refers to the whole phenomenon --- i.e., the \textit{entanglement} of object and apparatus, with all of the latter's attendant techniques --- rather than passively representing an object of measurement.

These changes, coupled with the apparent inability to generate a clean, mess-free signal under any circumstances, challenge the very premise that EMG represents an underlying state of the body which would be present in identical form whether or not it was measured. \hl{In designing with EMG, and beyond to other instances of biofeedback, we inherently change the original mechanic. Research that involves cases wherein the user is aware of such feedback must acknowledge that the originally observed phenomena ceases to exist through this practice.} Sensory feedback from EMG systems tends to defamiliarise one's own body, which can be a powerful design tactic \cite{wilde2017embodied, loke-movingmakingstrange}; this defamiliarisation shares common characteristics with soma design, for example through \textit{sensory misalignment} \cite{tennent2020soma}. 
\hl{As in the way we negotiate the messiness, we must also acknowledge that EMG is in fact part of the mechanic itself and make clear the intention, tactics, and goals of such designs.}

\subsection{Transparency and the Terminus of Experience}

\hl{Moving from concerns of materiality to intentionality,} the shift of awareness that occurs as the body is instrumentalised in relation to EMG can be understood phenomenologically from the perspective of Heidegger's \textit{zuhanden} and \textit{vorhanden} dichotomy, which marks an reversal of interface transparency as the instrument reappears to the performer, moving from a tool that is ``ready-to-hand'' towards one that is ``present-at-hand''. This framework for focusing on the role of attention when using an interactive system is extended to HCI in Dourish's embodied interaction approach \cite{dourish2001action}, allowing for a perspective on EMG that extends beyond binaries of matter and representation, and beyond the Cartesian dualism of split agendas between body and mind, to include a hybrid entity of human-machine relations. This approach is especially apt for framing EMG interactions, in which there is no longer any obvious external tool ``at hand''; instead, the tool \textit{becomes} the hand.  

Ihde's concept of the \textit{terminus} of experience \cite{ihde1990technology} offers a further phenomenological refinement of the role of the EMG apparatus. Where is the performer's attention and intention directed when wearing sonified EMG electrodes? Not on an abstract idea of intentionality itself: the performer must direct their focus somewhere specific and perceptible. In a kind of inversion of Merleau-Ponty's oft-cited example where a blind person learns to perceive the world through the tip of a cane \cite{merleau1962phenomenology}, perhaps EMG interfaces might draw the performer's attention inward to the internal bodily machinery itself. At question is whether we focus on the apparatus itself, or whether it recedes into the background as we carry out an interaction.  \hl{Independent of digital technology,} musicians will often describe the experience of clumsily learning an instrument until muscle memory and mastery enable them to become but the conduit through which the music speaks. But this perceived transparency of traditional musical instruments is an effect that emerges over years of learning \cite{nijs2017incorporation} as performers internalise a specialised set of \textit{bodily} and \textit{cultural techniques} \cite{doi:10.1080/03085147300000003, siegert2013cultural}. Likewise, in the case of EMG, at least from a sensorimotor skill perspective, playing with audible neural impulses and shifting the terminus of experience will be as unfamiliar as picking up a new musical instrument for the first time. Intentionality is thus enacted as a hybrid product of the (uninstrumented) human, the specific EMG apparatus (including all of its choices on signal processing and sonification strategies), and a cultivated set of playing techniques for bringing the two together. \hl{Appeals to a logic of transparency that would obfuscate these multiple stages of mediation are, from this vantage, being misled by what some have critiqued as "a dream of moving outside representation understood as bias and distortion," and by a corresponding effort to "uncover the true essence" of a technological system} \cite [p.77]{https://doi.org/10.1111/comt.12052}.


\subsection{EMG as Boundary-Making Practice}

From an agential realist perspective, the boundaries between humans, things and cultural systems are not fixed but enacted through material-discursive practice \cite{barad2003posthumanist}. Our EMG systems enact or reconfigure several boundaries:

\begin{itemize}
    \item \textbf{Between mind and body:} As discussed in Section 2.2, EMG signals defy Cartesian dualism, as they are not easily attributed to either the mind or the body as separate entities.
    \item \textbf{Between human body and electrical system:} The discourse around EMG supposes that neural impulses are just another form of electricity, completing a circuit with the apparatus of electrodes and amplifiers. This electrical view is only a partial understanding of what motor neurons are doing, and the impossibility of measuring their electrical signalling without changing the mind-body processes that produce those signals in the first place creates a fluid boundary between human and technical system.
    \item \textbf{Between technical components:} Our case studies show that the whole technical apparatus matters: electrodes, analog circuits, signal processing, mapping, sound production all co-determine the character of the result. No decision is neutral: even basic amplitude following changes the character of the sonic response and hence the interactions that emerge.
    \item \textbf{Between signals and concepts:} Some of our case studies used EMG in the context of modular synthesis, which operates on a control voltage (CV) paradigm where electrical signals are neatly mapped to stable musical concepts like pitch or amplitude. A particularly interesting effect that emerges with using EMG signals as CVs is that its complex stochastic character destabilises the relationship between signal and concept. EMG applied to an amplitude CV could produce effects of texture or pitch; applying it to a pitch CV could produce effects of timbre, demonstrating the underlying fragility and contingency of the signal-concept relationship to begin with.
    \item \textbf{Between artistic practices:} Some of our case studies explore EMG overlaid on existing instrumental or vocal practices. Performers adopt different strategies to navigate between familiar and novel practices, which include switching attention over time between traditional practice and EMG, using different muscle groups to separate the practices in space, or to deliberately overlay the two practices and embrace the messiness and ambiguity of the result.
\end{itemize}

\subsection{Designing with Movable Boundaries}

\hl{Looking across our case studies and the EMG-related HCI literature, we find persistent tensions and misconceptions around what EMG can do, what we as designers might \textit{want} it to do, and the sometimes perplexing experience of how specific EMG systems actually behave. This can be a source of friction for users, as in Section 3.3.3 where vocalists wanted the EMG to tell them about the quality and correctness of their vocal technique. On the other hand, the tension between expectations and observations can become a source of creative ideas} \cite{Benford2020:control}.

\hl{Both the frustrations and the creative opportunities of EMG stem in part from the way it challenges boundaries and familiar dualisms, as we enumerated in the previous section. Furthermore, design decisions are tightly entangled with aesthetics, contexts of use, and discursive elements, such as on-boarding instructions given to users and the ways outcomes are rationalised in publications and artistic programme notes. The challenge is that many of these intra-acting elements cannot be known at the outset of a design process or will change partway through. How then might a designer chart an initial course in a landscape of movable boundaries?

A possible mindset could be to maintain a metaphorical ``relaxed grip'' on concepts and representations in the design process. Athletes, musicians and craftspeople are often taught to keep a relaxed posture in relation to their tools and instruments, making it easier to perceive and react to the world through those tools. In Section} \ref{sec:representationalism} \hl{we critique rigid representationalism in working with EMG, but representations and conceptual frames are still useful ways of making sense of the world. Designing with a relaxed grip could entail starting with particular intended relationships between signals and concepts while maintaining an open mind to what else those signals might mean, and being ready to refine or replace those preconceptions when the momentum of the design process renders the original meaning untenable. 

The designer thus seeks to, in Ingold's words, ``find the grain of the world's becoming and to follow its course while bending it to their evolving purpose''} \cite{ingold2010textility}. \hl{The relaxed grip neither seeks to force technical systems and users to conform to pre-established boundaries, nor settles into aimless wandering. A challenge of this approach is knowing when to hold tight to a concept and when to let it slip (perhaps one of the ways that design, of DMIs at least, can be as much art as science} \cite{jorda2004instruments}). \hl{Another challenge is maintaining the same mindset throughout the entire deployment process, for example in giving instructions to users on the meaning of a system, allowing a degree of openness to reconfiguration by users, or in the avoidance of oversimplified post hoc narratives of what was designed and why.

We have endeavoured in the hardware and software we create for EMG systems and other DMIs to support designers to create interactive systems where these boundaries can be flexible. The question of volitional control or sonifying states of being is not a binary, developmental choice in Case Study 1, but a permanent foil counterposing intention and intuition. We hope that the discourse presented here introduces future designers wishing to work with the EMG to these questions. With this in mind, Fierro and Tanaka's modular EMG software suite} \cite{dimaggio_AM2023} \hl{is designed in such a way so as to support designers entering the field to work with these moving boundaries, from unprocessed EMG to feature extraction, from audification through parametric control to neural network regression. Likewise, the notion of `hackable instruments'} \cite{zappi2018hackable} \hl{extends this beyond the special case of the EMG to open up DMI instrument building and performance to accommodate hacking and modification, loosening up the instrument or interactive system from determinist top-down design.  
}

\section{Conclusion}
Returning to the question of why EMG researchers come and go, it is useful to consider once more the case of AlterEgo \cite{kapur2018}, the device that inspired the vocal EMG case study \cite{Reed_NIME20_VocalsEMG}, and which has all but disappeared since the final update of the project in 2020\footnote{https://www.media.mit.edu/projects/alterego/updates/}. After its initial release in 2018, AlterEgo garnered massive attention via TED talks and various technology reviews in outlets such as Popular Science \cite{PopularScience_AlterEgo}. What's more, the device was presented at a NeurIPS workshop as having medical promise within the context of Multiple Sclerosis-associated dysphonia \cite{kapur:alterego-ms}; but since then, we've heard nothing. It is interesting (and perhaps discouraging) to witness the abandonment of a device that had received so much praise for its potential to dramatically change human-computer interaction. The same is true of many other EMG devices explored within HCI; for instance, others presented at ACM or tangential conferences, such as Monarch \cite{Hartman:2015}, RAW \cite{Erdem20}, and MuscleSense \cite{Lim:2020}, which demonstrated the wide applicability of EMG in facilitating a variety of communicative, artistic, and therapeutic interactions. These EMG projects have either disappeared or gravitated towards other biosignal inputs, and their sudden demise has not been due to a lack of technical refinement; all of the aforementioned examples demonstrate careful approaches to signal filtering and mapping. So what is it? Are some applications just doomed to transient novelty status, with no staying power? Or is there an overarching reason why researchers abandon the EMG after short bursts of investigation? 

Perhaps the key to enduring EMG systems lies in this understanding of what the signal really is --- some messy electricity passing across muscle fibres, skin, and sensors, and \textit{not} a representation of any kind of truth or transparency. \hl{But this relational approach to understanding messy entanglements requires relinquishing certain ideas about instrumental control, and it thus sits in direct opposition to the kinds of exaggerated rhetoric that is used to market commercial EMG applications. One sees this in the recent reappearance of the Myo armband, which is now being developed by Meta as an EMG controller to be used with virtual and augmented reality systems. When Meta acquired CTRL-Labs --- the maker of the original Myo band --- the apparent motive was to build more ``natural, intuitive ways to interact with devices and technology'' }\footnote{https://www.facebook.com/1681/posts/10109385805377581}, \hl{and in a 2021 Meta Reality Labs promo video, the beta version of a new Meta EMG band is advertised as being able to grant users the ``purest form of a superpower''.} \footnote{https://youtu.be/hc2ADtNgeCM?feature=shared} 
\hl{The simultaneous naturalisation and deification of EMG in this example speaks not so much to a posthumanist concern with the entanglement between humans and nonhumans, and neither with the ``ethico-onto-epistemology'' of such relations} \cite{barad2007meeting}, \hl{but rather to a thinly veiled desire for transhumanism. The difference may seem subtle, but the latter signifies a problematic push toward narratives of human enhancement via technosolutionism, whereas the posthumanist approaches that interest us here largely come from research in feminist science studies and are concerned with the philosophical (and political) question of how Western, Enlightenment-era classifications of the human as an autonomous subject are being challenged by critical theoretical orientations and (digital) information technologies. Along with this, for us, comes a consideration of how ``more-than-human'' ontologies can shed light on the interconnectedness of materials and representations in EMG music systems, and more broadly, of how such a relational approach might contribute to articulating a non-reductive theory of entangled design} \cite{morrison_chi2024}. 

\hl{We recognise the importance of keeping theoretical reflection grounded in empirical work, and not letting theories become a one-size-fits-all template for analysing any and all technocultural phenomena, no matter how distinct. But the case of EMG provides an especially fitting illustration of how diverse elements, human and otherwise, come to matter and are mutually constituted in the context of a measuring apparatus. Moreover, it provides a novel situation for rethinking traditional notions of instrumentality and control in artistic (and specifically, musically) applications. Rather than a means to some musical ends, this paper has shown how EMG instruments translate and reconfigure the very language and conceptual categories that performers use to make sense of their practice, drawing attention to reciprocal flows of determination between materiality and intentionality in musical interactions, and to the way these dynamic flows trouble overly rigid boundaries between bodies, signals, and representations.}

\begin{acks}
We would like to thank Lou Barnell and Robin Foster for their work in the vocal EMG explorations in \textit{To Fly Out in the Heat of Day}, which is produced with support from the Sound and Music UK \textit{New Voices} programme, Northern Lights Project, and Royal Birmingham Conservatoire as part of Lou's ongoing practice-based Research into `Live Dreaming'. LM and AM are funded by the UKRI Frontier Research grant EP/X023478/1 (`RUDIMENTS'). AT has received funding from the European Research Council  (FP/2007-2013) / ERC Grant FP7-283771 (`META GESTURE MUSIC') and from the Horizon 2020  grant  no. 789,825 (`BIOMUSICAL INSTRUMENT'). AT and DF have been supported by funding from the French Agence Nationale de la Recherche (ANR-21-CE38-0018), `Brain-Body Digital Musical Instrument (BBDMI).

\end{acks}

\bibliographystyle{ACM-Reference-Format}
\bibliography{DIS_2024}









\end{document}